\documentclass[ALICE,manyauthors]{cernphprep}

\usepackage[comma,square,numbers,sort&compress]{natbib}
\usepackage{hyperref}
\usepackage{lineno}
\usepackage{tikz}
\usepackage[section]{placeins}
\usepackage {multicol}
\usepackage {multirow}
\usepackage {units}
\usepackage {array}
\usepackage{verbatim}
\usepackage{dblfloatfix} 
\usepackage[T1]{fontenc}
\usepackage{orcidlink}

\newcommand{\pt}           {\ensuremath{p_\mathrm{T}}}

\newcommand{\pttrig}{\ensuremath{p_\mathrm{T,\,trig}}}
\newcommand{\ptassoc}{\ensuremath{p_\mathrm{T,\,assoc}}}
\newcommand{\ptjet}{\ensuremath{p_\mathrm{T,\,jet}^\mathrm{ch}}}
\newcommand{\ptlead}{\ensuremath{p_\mathrm{T,\,LP}}}

\begin{document}

\begin{titlepage}

\PHyear{2023}
\PHnumber{195}      
\PHdate{30 August}  
%

\title{Multiplicity and event-scale dependent flow and jet fragmentation in pp collisions at ${\sqrt{{\textit s}}}=\mathbf{13}$~TeV and in p--Pb collisions at ${\sqrt{{\textit s_\mathrm{NN}}}} = \mathbf{5.02}$~TeV}
\ShortTitle{Flow and jets in pp and p--Pb collisions}   

\Collaboration{ALICE Collaboration\thanks{See Appendix~\ref{app:collab} for the list of collaboration members}}
\ShortAuthor{ALICE Collaboration} 

\begin{abstract}
Long- and short-range correlations for pairs of charged particles are studied via two-particle angular correlations in pp collisions at ${\sqrt{{\textit s}}}=13$~TeV and p--Pb collisions at ${\sqrt{s_\mathrm{NN}}} = 5.02$~TeV. The correlation functions are measured as a function of relative azimuthal angle $\Delta\varphi$ and pseudorapidity separation $\Delta\eta$ for pairs of primary charged particles within the pseudorapidity interval $|\eta| < 0.9$ and the transverse-momentum interval 1~$ < \pt < $~4~GeV/$c$. Flow coefficients are extracted for the long-range correlations ($1.6 < |\Delta\eta| <1.8$) in various high-multiplicity event classes using the low-multiplicity template fit method. The method is used to subtract the enhanced yield of away-side jet fragments in high-multiplicity events. These results show decreasing flow signals toward lower multiplicity events. Furthermore, the flow coefficients for events with hard probes, such as jets or leading particles, do not exhibit any significant changes compared to those obtained from high-multiplicity events without any specific event selection criteria. The results are compared with hydrodynamic-model calculations, and it is found that a better understanding of the initial conditions is necessary to describe the results, particularly for low-multiplicity events.

\end{abstract}

\end{titlepage}

\setcounter{page}{2}

\section{Introduction}
\label{sec:intro}

High-energy nucleus--nucleus (AA) collisions exhibit strong collectivity, which has been observed through anisotropy in the momentum distribution of emitted final-state particles at RHIC~\cite{Adams:2005dq,Adcox:2004mh,Arsene:2004fa,Back:2004je} and the LHC~\cite{Abelev:2012di, Abelev:2014pua, ATLAS:2011ah,ALICE:2022wpn}. This momentum anisotropy is developed by the pressure-driven expansion of the strongly interacting quark--gluon plasma (QGP), which emerges from the initial spatial anisotropy in such collisions.
The collective nature of the momentum anisotropy is mostly deduced via particle correlations which span over a wide range of pseudorapidity. The collective motion of the emitted particles, which reflects the collectivity of the initial medium, is generally quantified using a Fourier expansion, characterizing the so-called ``anisotropic flow''~\cite{Ollitrault:1992bk}. In recent years, long-range correlations have been also observed in smaller collision systems such as high-multiplicity proton--proton (pp)~\cite{ATLAS:2015hzw,Khachatryan:2015lva,Khachatryan:2016txc,Acharya:2019vdf,ATLAS:2017rtr,CMS:2010ifv,LHCb:2015coe}, proton--nucleus (pA)~\cite{ALICE:2012eyl,ATLAS:2014qaj,ATLAS:2016yzd,Khachatryan:2016ibd}, and in collisions of light nuclei~\cite{PHENIX:2018lia,Aidala:2017ajz}. 
These observations raise the question to what extent do small-system collisions and heavy-ion collisions share the underlying mechanism, which is responsible for the observed long-range correlations.
A crucial evidence of a strongly interacting medium in small-system collisions would be the presence of jet quenching~\cite{Gyulassy:1990ye,Wang:1991xy}. However, this phenomenon has not yet been observed in either high-multiplicity pp or p--Pb collisions~\cite{Adam:2014qja,Khachatryan:2016odn,Adam:2016jfp,Adam:2016dau,Acharya:2017okq}, possibly due to the current experimental uncertainties being too large to observe it in such small-system collisions.

Current approaches to model heavy-ion collisions divide the evolution of the out-of-equilibrium, strongly-coupled, quantum-chromodynamic medium into multiple stages, and each stage is described by an effective theory. To this date, the combination of color-glass condensate effective field theory (CGC-EFT)~\cite{Schenke:2012wb,Schenke:2012hg}, causal hydrodynamics~\cite{Kolb:2003dz,Song:2007ux,Dusling:2007gi,Holopainen:2010gz,Schenke:2010rr,Romatschke:2007mq,Niemi:2015qia,Jeon:2015dfa,Romatschke:2017ejr}, and a hadronic cascade model~\cite{Bass:1998ca,Bleicher:1999xi,Weil:2016zrk} leads to the most successful description of a wide range of observables in heavy-ion collisions, e.g., particle spectra, centrality dependence of average particle transverse momenta, and multi-particle correlations~\cite{ALICE:2016kpq,Acharya:2017gsw,Acharya:2017zfg,Acharya:2020taj,ALICE:2021klf,ALICE:2021adw,ALICE:2013mez,ALICE:2011ab}. 
By employing global Bayesian analyses, parameters of the multi-stage model, including those quantifying the transport properties of the QGP, can be constrained using measured data~\cite{Bernhard:2016tnd,Bernhard:2019bmu,Parkkila:2021tqq,Parkkila:2021yha}.
Despite the studies describing both heavy AA and pA collisions in a single framework~\cite{Moreland:2018gsh}, the origin of the flow-like correlations is still under debate. It is unclear whether the flow-like behavior originates from the early stages of the collision in the realm of applicability of CGC-EFT~\cite{Dusling:2012cg,Bzdak:2013zma} or whether it develops during the collective evolution, where causal hydrodynamics is applicable~\cite{Greif:2017bnr,Mantysaari:2017cni}. Both scenarios may be responsible for the observed correlations in the final state~\cite{Greif:2017bnr}. Although collective models are successful in describing available two-particle correlation data from small-system collisions, they predict the opposite sign for four-particle azimuthal cumulants compared to experiment~\cite{Khachatryan:2016txc,ATLAS:2017rtr,Zhao:2017rgg}. On the other hand, a semi-analytical toy model based on the Gubser hydrodynamic solution~\cite{Gubser:2010ze,Gubser:2010ui} can explain the two- and four-particle correlations in pp collisions~\cite{Taghavi:2019mqz}. In particular, this model has explained the relationship between the sign of the four-particle cumulants and fluctuations in the initial state~\cite{Taghavi:2019mqz}. 

Besides the models based on the causal hydrodynamic framework, there are other attempts to explain the observed flow-like signals in small-system collisions using alternative descriptions. 
For instance, a study based on the A Multi-Phase Transport model (AMPT)~\cite{Lin:2004en} leads to satisfactory agreement with the experimental data~\cite{OrjuelaKoop:2015jss}. The applicability of fluid-dynamical simulations and partonic cascade models in small-system collisions was explored in Ref.~\cite{Gallmeister:2018mcn}. In a kinetic-theory framework with isotropization-time approximation, it is possible to explain the long-range correlations by fluid-like (hydrodynamic) excitation for Pb--Pb collisions and particle-like (or non-hydrodynamic) excitation for pp or p--Pb collisions~\cite{Kurkela:2019kip,Kurkela:2020wwb,Ambrus:2021fej}. Another potential description for the collectivity in small-system collisions is provided by PYTHIA~8, in which interacting strings repel one another in a transverse direction by a mechanism dubbed as ``string shoving''~\cite{Bierlich:2017vhg,Bierlich:2019ixq}. The repulsion of the strings causes microscopic transverse pressure, giving rise to long-range correlations of particles. The string shoving approach in PYTHIA~8 successfully reproduces the near-side ridge yield observed in measurements by ALICE~\cite{ALICE:2021nir} and CMS~\cite{Khachatryan:2016txc}. A systematic mapping of correlation effects across collision systems of various sizes is currently underway on the theoretical side, for example, see Ref.~\cite{Schenke:2020mbo}. A quantitative description of the full set of experimental data has not been achieved yet. A summary of various explanations for the observed correlations in small-system collisions is given in Refs.~\cite{Strickland:2018exs,Loizides:2016tew,Nagle:2018nvi}. 

Measurements of anisotropic flow in small-system collisions are strongly affected by non-flow effects, predominantly originating from correlations among the constituents of jet fragmentation processes.
In case of two-particle correlations, the non-flow contribution is usually suppressed by requiring a large $\Delta\eta$ gap between the two particles. This separation in pseudorapidity is also widely used in cumulant methods~\cite{Bilandzic:2010jr, Acharya:2019vdf}. However, this $\Delta\eta$-gap method removes the non-flow contribution only on the near side ($\Delta\varphi\sim0$) and not on the away side~($\Delta\varphi\sim\pi$). Later, a low-multiplicity template fit method was proposed to remove non-flow contributions on the away-side~\cite{ATLAS:2015hzw, ATLAS:2016yzd, ATLAS:2018ngv}. This method takes into account that the yield of jet fragments increases with increasing particle multiplicity~\cite{CMS:2013ycn, ALICE:2013tla, ALICE:2014mas}.
By using the template fit method, the yield of away-side jet fragments can be subtracted, provided that the distribution that quantifies the shape of jet fragments is independent of the multiplicity class and therefore can be described by the low-multiplicity template.

As an extension of the studies of the near-side long-range ridge and jet-fragmentation yields in pp collisions at the center-of-mass energy $\sqrt{s}=13$ TeV~\cite{ALICE:2021nir} and in p--Pb collisions at the center-of-mass energy per nucleon pair $\sqrt{s_{\mathrm{NN}}}=5.02$ TeV~\cite{ALICE:2012eyl,ALICE:2013snk}, this article studies the interplay of jet production and collective effects, i.e., short- and long-range correlations simultaneously in these systems. The article also reports flow coefficients extracted for collisions tagged with different event-scale selections. The event-scale selection requires a minimum transverse momentum of the leading particle or the reconstructed jet at midrapidity, which is expected to bias the impact parameter of pp collisions to be smaller on average~\cite{Sjostrand:1986ep,Frankfurt:2003td,Frankfurt:2010ea}. At the same time, the transverse momentum of the leading particle or the reconstructed jet provides a measure of the four-momentum transfer ($Q^2$) in the hard-parton scattering~\cite{Chatrchyan:2012tt, Chatrchyan:2011id,ALICE:2022fnb}. The transverse-momentum threshold implies a higher $Q^2$ for the collision. Such events with a large $Q^2$ may, on average, have a lower impact parameter than pp events without any requirement on $Q^2$~\cite{Frankfurt:2003td}.

This article is organized as follows. First, the experimental setup and analysis method are described in Sec.~\ref{sec:experiment} and Sec.~\ref{sec:ana}, respectively. Section~\ref{sec:uncertainties} discusses the systematic uncertainties. The results and their comparison with model calculations are presented and discussed in Sec.~\ref{sec:results}. Finally, the results are summarized in~Sec.~\ref{sec:summary}.

\section{Experimental setup and data samples}
\label{sec:experiment}

The analysis is based on pp and p--Pb data collected during the LHC Run 2 period. The pp collisions had a center-of-mass energy $\sqrt{s} = 13$~TeV, and they were recorded from 2016 to 2018.
The p--Pb collisions had a center-of-mass energy per nucleon--nucleon pair $\sqrt{s_\mathrm{NN}} = 5.02$~TeV, and they were collected in 2016. It is worth noting that in p--Pb collisions there is a shift in the center-of-mass rapidity of $\Delta y = 0.465$ in the direction of the proton beam due to the asymmetric collision system.

A comprehensive description of the ALICE detector and its performance can be found in Refs.~\cite{Aamodt:2008zz, Abelev:2014ffa, ALICE:2022wpn}. The analysis utilizes the V0 detector~\cite{Abbas:2013taa}, the Inner Tracking System (ITS)~\cite{aliceITS, ALICE:2013nwm}, and the Time Projection Chamber (TPC)~\cite{aliceTPC}. 

The V0 detector consists of two stations on both sides of the interaction point, V0A and V0C, each comprising 32 plastic scintillator tiles, covering the full azimuthal angle within the pseudorapidity intervals $2.8 < \eta < 5.1$ and $-3.7 < \eta < -1.7$, respectively. 
The ITS is a silicon tracker with six layers of silicon sensors. The two innermost layers of the ITS are called the Silicon Pixel Detector (SPD)~\cite{Santoro2009:ALICESPD}. In addition to the two SPD layers, the middle two layers are the Silicon Drift Detector, and the outermost layers are the Silicon Strip Detector. The TPC is a gas-filled cylindrical tracking detector providing up to 159 reconstruction points for charged tracks traversing the full radial extent of the detector.

The V0 provides a minimum bias (MB) trigger in both pp and p--Pb collisions and an additional high-multiplicity trigger in pp collisions. The MB trigger is obtained by a time coincidence of V0A and V0C signals. 
Amplitudes of V0A and V0C signals are proportional to charged-particle multiplicity, and their sum is denoted as V0M.  
The high-multiplicity trigger in pp collisions requires the V0M signal to exceed five times the mean value measured in MB collisions, selecting the 0.1\% of MB events with the largest V0M multiplicity. The centrality in p--Pb collisions is determined using the V0A detector, which is located in the Pb-going direction~\cite{Adam:2014qja}. The analyzed data samples of MB and high-multiplicity pp events at $\sqrt{s}=13$~TeV correspond to integrated luminosities ($\mathcal{L}_\mathrm{int}$) of about 19 nb$^{-1}$ and 11 pb$^{-1}$, respectively~\cite{ALICE-PUBLIC-2016-002}. In p--Pb collisions at $\sqrt{s_\mathrm{NN}} = 5.02$~TeV, the corresponding integrated luminosity is $\mathcal{L}_\mathrm{int} \sim 0.3$ nb$^{-1}$. 

Positions of primary vertices are reconstructed from signals measured by the SPD. The reconstructed primary vertices are required to be within 8 cm of the nominal interaction point along the beam direction. 
Pileup events are identified as events with multiple reconstructed primary vertices. These events are rejected if the distance between any of the vertices to the main primary vertex is greater than 0.8 cm.
The probability of pileup events is estimated to range from $10^{-3}$ to $10^{-2}$ for MB and high-multiplicity events in pp collisions~\cite{ALICE:2020swj}. The pileup probability is estimated to be negligible in p--Pb collisions~\cite{Acharya:2017okq}. 

Charged-particle tracks are reconstructed using the combined information from the ITS and TPC.
For charged particles emitted from a vertex located within $|z_\mathrm{vtx}|<8$ cm along the beam direction, the ITS and TPC provide a pseudorapidity coverage of $|\eta|<1.4$ and 0.9, respectively. Both detectors have full coverage in azimuth. They are placed in a uniform magnetic field of 0.5 T that is oriented along the beam direction.

The charged-particle selection criteria are optimized to ensure a uniform efficiency over the midrapidity range $|\eta|<0.9$ to mitigate the effects of small areas where some ITS layers are inactive in both collision systems. The selected sample of tracks consists of two classes. Tracks in the first class must have at least one hit in the SPD. Tracks of the second class do not have any hits in the SPD, but their origin is constrained to the primary vertex~\cite{ALICE:2012eyl}. 
Charged-particle tracks are reconstructed down to a transverse momentum ($\pt$) of 0.15 GeV/$c$ with an efficiency of approximately 65\%~\cite{Ivanov:2006yra}. The efficiency increases to 80\% for particles with $\pt>1$ GeV/$c$. The $\pt$ resolution is approximately 1\% for primary charged particles~\cite{ALICE-PUBLIC-2017-005} with $\pt<$ 1~GeV/$c$, and it linearly increases to 6\% at $\pt\sim$ 50~GeV/$c$ in pp collisions and 10\% in p--Pb collisions~\cite{ALICE:2018vuu}.

\section{Analysis procedure}
\label{sec:ana}
\subsection{Two-particle angular correlations}
Two-particle angular correlations are measured as a function of the relative azimuthal angle ($\Delta\varphi$) and the relative pseudorapidity ($\Delta\eta$) between a trigger and associated particles
\begin{eqnarray}
\frac{1}{N_{\rm{trig}}} \frac{ \rm{d}\it{}^{2} N_{\rm{pair}} }{ \rm{d} \Delta\eta \rm{d}\Delta\varphi} = B(0, 0)\frac{S(\Delta\eta, \Delta\varphi)}{B(\Delta\eta, \Delta\varphi)}  \Big\lvert_{\pttrig,\,\ptassoc},
\label{eq:corrfunction}
\end{eqnarray}
where $p_\mathrm{T,trig}$ and $p_\mathrm{T,assoc}$ denote the transverse momentum of the trigger and associated particles, respectively.
While the transverse momentum range for associated particles is fixed to $1<p_\mathrm{T,assoc}<4$~GeV/$c$ for trigger particles, several transverse momentum ranges are considered. The lower limit of $p_\mathrm{T,trig}$ and $p_\mathrm{T,assoc}$ ($>$~1~GeV/$c$) is chosen in order to avoid jet-like contributions from lower $p_\mathrm{T}$ particles which extend into the larger $\Delta\eta$ range because of the limited $\eta$ acceptance~\cite{ALICE:2021nir}. The numbers of trigger particles and trigger--associated particle pairs are denoted as $N_\mathrm{trig}$ and $N_\mathrm{pair}$, respectively.
The average number of pairs in the same event, denoted by $S(\Delta\eta, \Delta\varphi)$, is given by $\frac{1}{N_\mathrm{trig}}\frac{\mathrm{d^2}N_\mathrm{same}}{\mathrm{d}\Delta\eta\mathrm{d}\Delta\varphi}$. The $B(\Delta\eta, \Delta\varphi)$ represents the number of pairs in mixed events and is normalized with its value at the point where $\Delta\eta=0$ and $\Delta\varphi=0$, denoted as $B(0,0)$. To correct for acceptance effects, $S(\Delta\eta, \Delta\varphi)$ is divided by $B(\Delta\eta, \Delta\varphi)/B(0,0)$. The particles are weighted by the inverse of the tracking efficiency, which is obtained in the same way as in Ref.~\cite{ALICE:2021nir}. In that study, the tracking efficiency and the secondary contamination (fake rate) were calculated using a detector simulation with the PYTHIA 8 event generator and the GEANT3 transport code~\cite{Brun:1994aa}. To account for differences in particle composition between real data and PYTHIA, the tracking efficiency is determined from the above mentioned PYTHIA-based simulation with reweighted primary particle-species composition.
The weights reflect realistic abundances of different particle species, which were extracted by a data-driven method~\cite{ALICE:2018hza, ALICE:2018vuu}.
Events to be mixed are required to have primary vertices within the same 2 cm wide $z_{\rm vtx}$ interval. The correlation functions are averaged over the vertex intervals, resulting in the final per-trigger yield~\cite{Kopylov:1974th,Adam:2016tsv}. 

\begin{figure}[h!]
		\includegraphics[width=0.5 \textwidth]{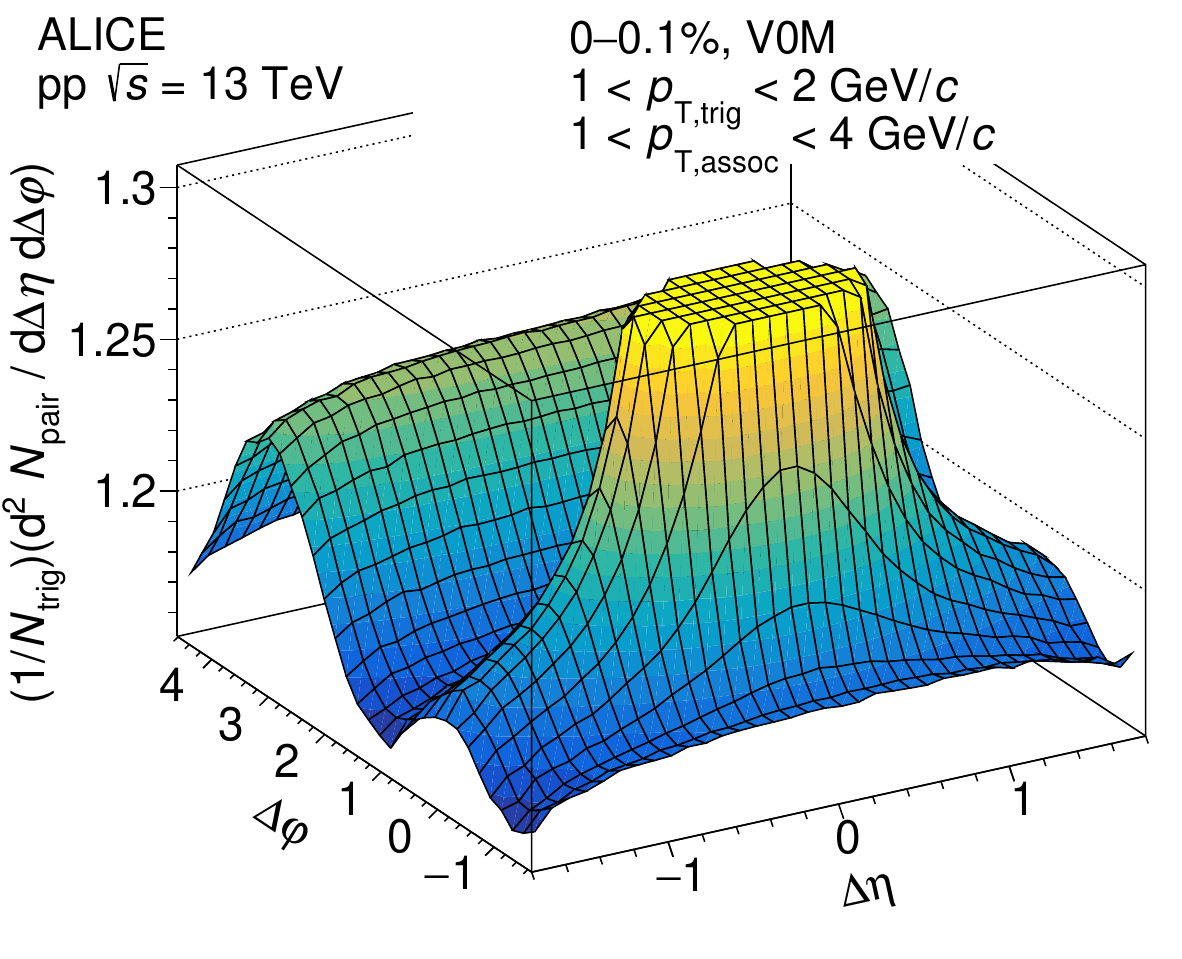} 
		\includegraphics[width=0.5 \textwidth]{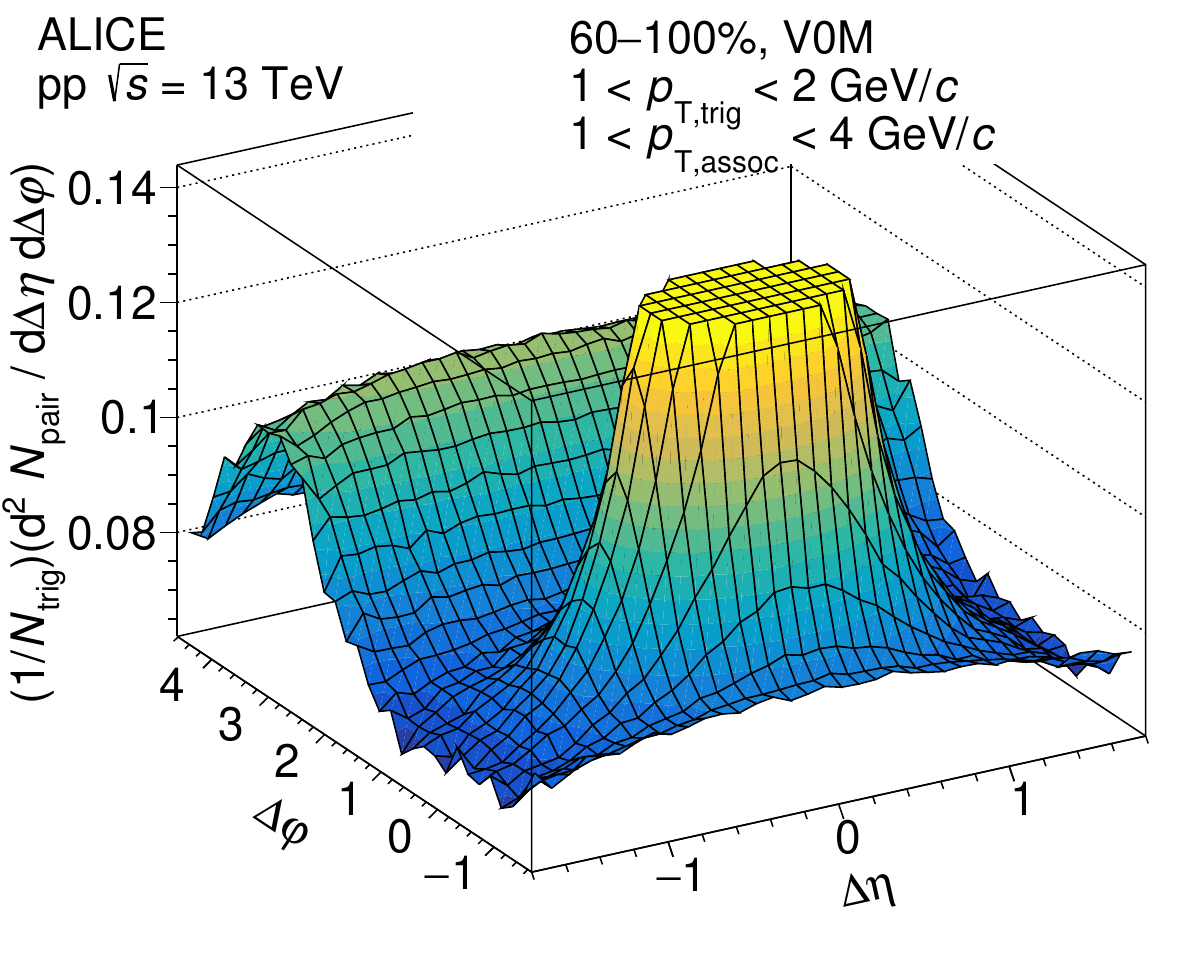} 
  		\includegraphics[width=0.5 \textwidth]{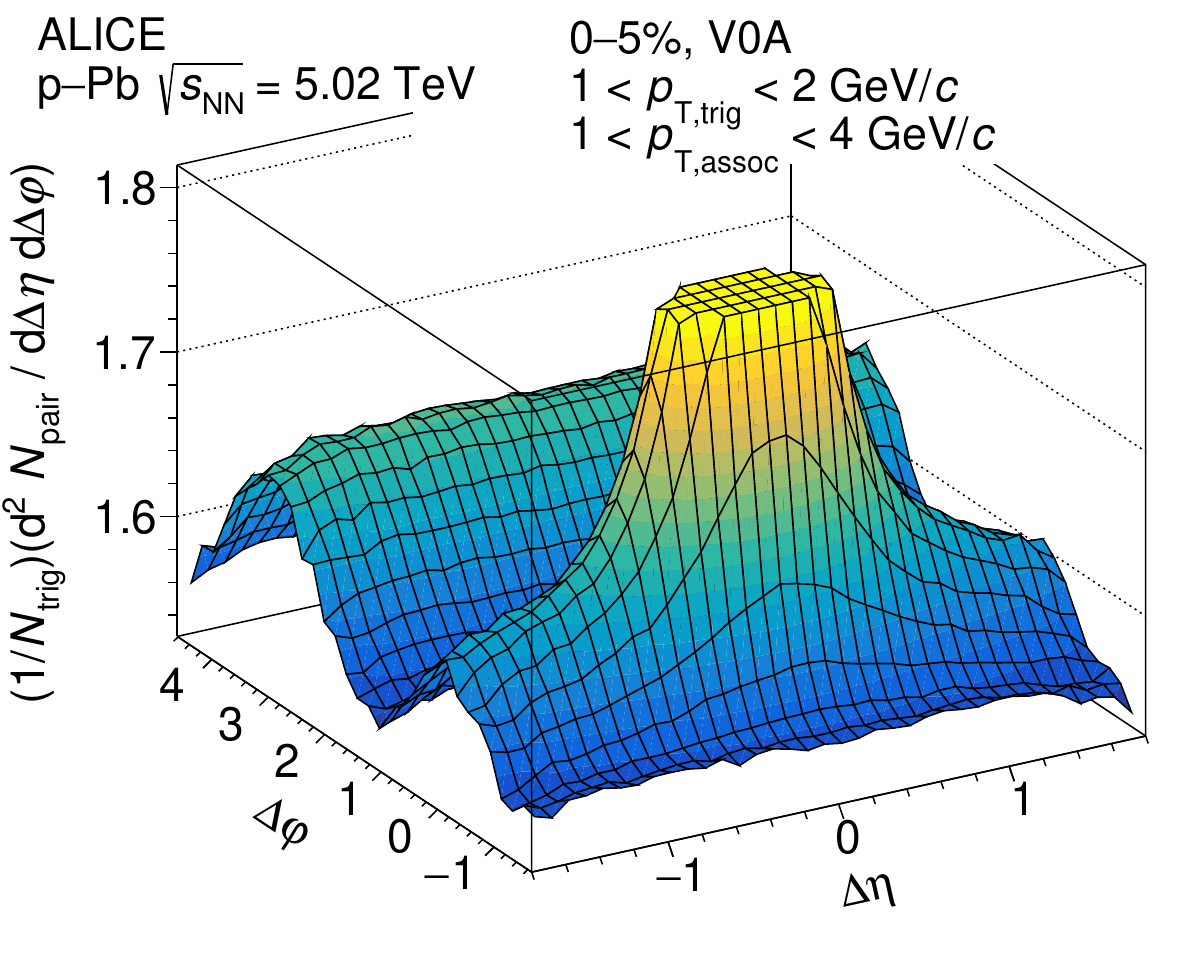}
		\includegraphics[width=0.5 \textwidth]{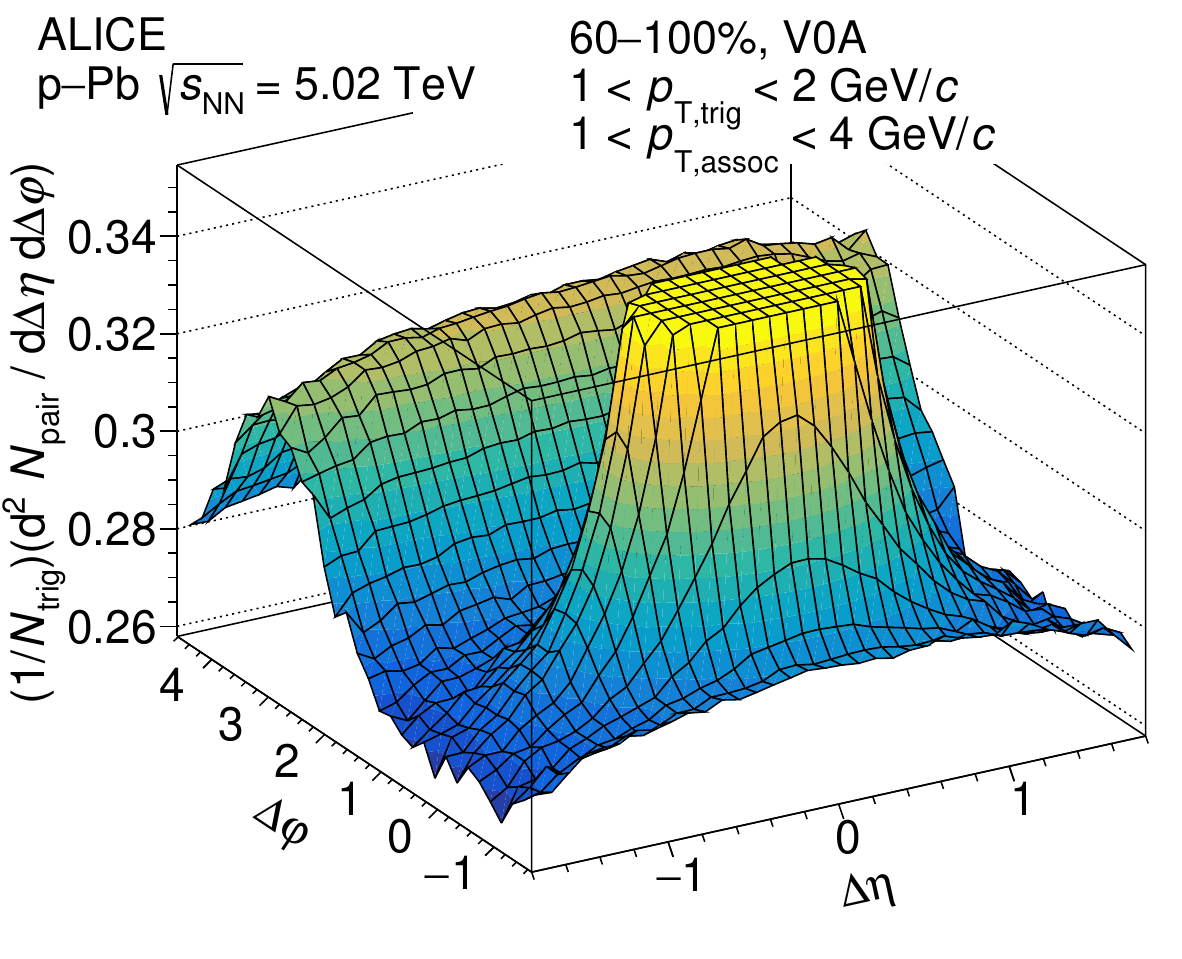}
\caption{Two-dimensional correlation functions are presented for high-multiplicity (0--0.1\% or 0--5\%, on the left) and low-multiplicity (60--100\%, on the right) events in $\sqrt{s}=13$~TeV pp collisions in the top panels. The corresponding distributions for $\sqrt{s_{\mathrm{NN}}}=5.02$~TeV p--Pb collisions are shown in the bottom panels. All correlation functions are shown for 1~$<\it{p}_{\rm{T,trig}}<$~2~GeV/$c$ and 1~$<\it{p}_{\rm{T,assoc}}<$~4~GeV/$c$, respectively.}
\label{fig:doubleridge}
\end{figure}

The fully corrected correlation functions from pp and p--Pb collisions are shown in Fig.~\ref{fig:doubleridge}. 
The $z$-axis is scaled in order to exhibit the ridge structures at large $\Delta\eta$ regions. As a result, the jet peaks are sheared off in all figures. The flow modulation structure is clearly observed to emerge in the high-multiplicity collisions for both systems, while it is not seen in the low-multiplicity collisions. The away-side regions are populated mostly by back-to-back jet correlations. 

The per-trigger yield is determined by integrating the correlation function at large $\Delta\eta$ ($1.6<|\Delta\eta|<1.8$) to remove non-flow contributions from near-side jet fragments.
The per-trigger yield as a function of $\Delta\varphi$ is expressed as

\begin{eqnarray}
Y(\Delta\varphi) = \frac{1}{N_{\mathrm{trig}}} \frac{\mathrm{d} N_{\mathrm{pair}}}{\mathrm{d}\Delta\varphi} = \int_{1.6<|\Delta \eta|<1.8} \left[\frac{1}{N_{\mathrm{trig}}} \frac{\mathrm{d}^{2}N_{\mathrm{pair}}}{\mathrm{d}\Delta\eta \mathrm{d}\Delta\varphi}\right] \frac{1}{\delta_{\Delta\eta}} \mathrm{d}\Delta \eta,
\label{eq:pertrigger}
\end{eqnarray}

where the factor $\delta_{\Delta\eta}=$~0.4 normalizes the obtained per-trigger yield per unit of pseudorapidity.
The per-trigger yields are extracted for the considered $p_{\mathrm{T,trig}}$ and $p_{\mathrm{T,assoc}}$ intervals in several multiplicity classes: 0--0.1\%, 1--5\%, 5--20\%, 20--60\%, and 60--100\% in pp collisions, and 0--5\%, 5--10\%, 10--20\%, 20--40\%, 40--60\%, and 60--100\% in p--Pb collisions. The conversion of the measured forward event multiplicities to the charge-particle multiplicities $N_{\mathrm{ch}}$ at midrapidity ($|\eta|<0.5$) used in Sec.~\ref{sec:results} is based on Ref.~\cite{ALICE:2020swj}.

\subsection{Extraction of flow coefficients}

As discussed in Refs.~\cite{ATLAS:2015hzw,ATLAS:2016yzd}, the correlation function in a given multiplicity interval is fitted with 
\begin{eqnarray}
\label{eq:narray}
Y_{\rm{HM}}(\Delta\varphi) = G~(1 + 2v_{2,2}\cos(2\Delta\varphi) + 2v_{3,3}\cos(3\Delta\varphi)) + F~Y_{\rm{LM}}(\Delta\varphi),
\label{eq:tmpfit}
\end{eqnarray}
where $Y_{\rm{LM}}(\Delta\varphi)$ is the measured per-trigger yield from low-multiplicity events. The normalization factor for the first three Fourier terms, which parameterize the long-range, flow-like correlation, is denoted as $G$. The scale factor $F$ compensates for the increased yield of away-side-jet hadrons in the analyzed multiplicity class relative to the low-multiplicity template that corresponds to the 60--100\% class~\cite{ALICE:2013tla,ALICE:2014mas}.
The fit determines the scale factor $F$, pedestal $G$, and $v_{n,n}$ and is performed in various high-multiplicity classes as well as in different $p_\mathrm{T,trig}$ intervals. 
This method assumes that $Y_{\rm{LM}}$ does not contain a near-side-peak structure that would originate from jet fragmentation or a near-side ridge.
Furthermore it is assumed that the shape of the away-side-peak structure remains the same when changing the multiplicity class.
The first assumption is ensured using the selected low-multiplicity template which does not have a strong near-side-peak structure compared to the studied higher-multiplicity classes. The second assumption, which involves the modification of jet shapes, was tested by projecting the near-side jet peaks onto $\Delta\eta$. 
This modification of the jet shape is considered as one of the sources of systematic uncertainty and will be discussed in Sec.~\ref{sec:uncertainties}.

\begin{figure}[h!]
	\centering
	\hspace{-3em}\includegraphics[width=0.6\textwidth]{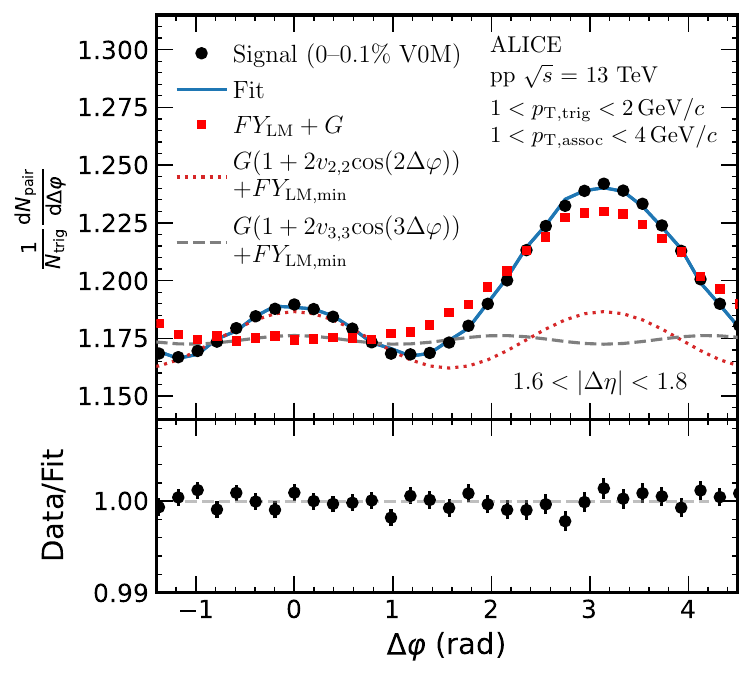} 
	\caption{Per-trigger yield in $1.6<|\Delta\eta|<1.8$ extracted from 0--0.1\% and 60--100\% multiplicity percentile events in $\sqrt{s}=13$~TeV pp collisions. The data are fitted with the template fit method described by Eq.~\ref{eq:tmpfit}. The black markers show the signal for the 0--0.1\% multiplicity percentile. The red squares correspond to the low-multiplicity signal. The red and gray curves correspond to the extracted $v_{2,2}$ and $v_{3,3}$ signals, respectively. To improve visibility, the baselines of flow signals are shifted by $FY_{\mathrm{LM,min}}$, which represents the minimum yield of $FY_{\mathrm{LM}}(\Delta\varphi)$. The signal-to-fit ratio is shown in the bottom panel. The $\chi^{2}$ divided by the number of degrees of freedom is 0.894.}
	\label{fig:flowext}
\end{figure}

Figure~\ref{fig:flowext} shows the template fit results for the 0--0.1\% multiplicity interval in pp collisions at $\sqrt{s} = 13$~TeV. Values of the extracted scale factor $F$ in different multiplicity intervals and systems are summarized in Table~\ref{tab:FpppPb}. In pp collisions, the value of $F$ is observed to increase slightly as the event multiplicity increases. The $F$ value, which is measured for the highest-multiplicity bin, is approximately 25\% larger than the value found for the 20--60\% bin.
A similar dependence on multiplicity is observed for p--Pb collisions, although the dependence on the multiplicity interval is weaker.
In the first three columns of Table~\ref{tab:FpppPb}, representing collisions with higher multiplicities, there is an increase in the $F$ value, while as shown by the subsequent columns, there is a decrease for lower multiplicity collisions.
When comparing the $F$ values from pp and p--Pb collisions, which have similar centrality, the value of $F$ in p--Pb collisions is found to be smaller and closer to unity.
This suggests that the jet fragmentation yield on the away-side increases with multiplicity, and that this feature is more pronounced in pp collisions. The difference between the two systems is likely to be explained by the true-geometry-driven centrality in p--Pb collisions, as opposed to the jet-dominated bias in pp collisions.
The previous analyses published by ALICE in Refs. ~\cite{ALICE:2012eyl,ALICE:2013snk} assumed that the jet contribution remains constant as a function of multiplicity (i.e. $F$ was assumed to be 1). However, this assumption may lead to an underestimation of non-flow contamination in the measurements of anisotropic flow.

\begin{table}[!b]
\caption{The scale factor $F$ for various multiplicity intervals in pp collisions (top) and p--Pb collisions (bottom), with 1~$<\it{p}_{\rm{T,trig}}<$~2~GeV/$c$ and 1~$<\it{p}_{\rm{T,assoc}}<$~4~GeV/$c$. The table reports statistical uncertainties only. Average systematic uncertainty of $F$ is about 3.8\% for both collision systems and multiplicity intervals.}
\centering
\resizebox{\textwidth}{!}{%
\begin{tabular}{|c|c|c|c|c|c|c|}
\hline
	V0M (pp)& 0--0.1\% & 1--5\% & 5--20\% & 20--60\% & &\\
\hline
	$F$ & 1.504$\pm$0.017 & 1.414$\pm$0.030 & 1.360$\pm$0.019 & 1.208$\pm$0.015 & & \\
\hline
V0A (p--Pb)& 0--5\% & 5--10\% & 10--20\% & 0--20\% & 20--40\% & 40--60\% \\
\hline
$F$& 1.135$\pm$0.026 & 1.140$\pm$0.026 & 1.152$\pm$0.021 & 1.145$\pm$0.017 &1.092$\pm$0.015 & 1.083$\pm$0.015 \\
\hline
\end{tabular}
}
\label{tab:FpppPb}
\end{table}

In the following, the near and away-side jet fragmentation yields are calculated to verify the template fit method by comparing the jet fragmentation yields to the PYTHIA model. The away-side jet fragmentation yields in the PYTHIA model are obtained using the standard $\Delta\varphi$ analysis~\cite{PHENIX:2006gto}, while in the data, the away-side jet fragmentation yields are extracted using the template fit method because of the flow modulations in the data. The comparison between the data and the PYTHIA model provides a validation of the template fit method. 

Equivalently to Eq.~\eqref{eq:pertrigger}, measured data are used to obtain the near-side $\Delta\eta$ correlations with
\begin{equation}
Y(\Delta\eta)=\frac{1}{N_{\rm{trig}}} \frac{ \rm{d}\it{}N_{\rm{pair}} }{ \rm{d}\Delta\eta }=\int_{|\Delta\varphi|<1.3}\left[ \frac{1}{\it{N}_{\rm{trig}}} \frac{ \rm{d}\it{}^{2} N_{\rm{pair}} }{ \rm{d}\Delta\eta d\Delta\varphi} \right]\dfrac{1}{\delta_{\Delta\varphi}} {\rm{d}}\Delta \varphi - D_\mathrm{ZYAM},
\label{eq:deltaetacorr}
\end{equation}
where $\delta_{\Delta\varphi}=2.6$ and $D_\mathrm{ZYAM}$ defines the baseline of the ZYAM background subtraction~\cite{Ajitanand:2005jj}. The baseline is obtained by finding the minimum of the distribution defined by the integral in Eq.~\eqref{eq:deltaetacorr}.
As flow has a weak $\eta$ dependence~\cite{ATLAS:2011ah,PHENIX:2018hho,ALICE:2016tlx}, the jet-fragmentation yield can be calculated after the ZYAM background subtraction~\cite{Ajitanand:2005jj}.
The near-side jet-like yields were extracted by integrating the $Y(\Delta\eta)$
\begin{eqnarray}
Y^\mathrm{near}_{\rm{frag}} = \int_{|\Delta \eta|<1.3} \left( \frac{1}{\it{N}_{\rm{trig}}} \frac{ \rm{d}\it{}N_{\rm{pair}} }{ \rm{d}\Delta\eta } \right) \rm{d} \Delta\eta.
\label{eq:Ynear}
\end{eqnarray}

The away-side jet-like yield in data is calculated by integrating the low-multiplicity template over $\pi/2<\Delta\varphi<3/2\pi$ and scaling it by the parameter $F$ from Eq.~\eqref{eq:narray}, $Y^{\rm{away, HM}}_{\mathrm{frag}} = Y^{\rm{away, LM}}_{\mathrm{frag}} \times F$. The $Y^{\rm{away, LM}}_{\mathrm{frag}}$ is directly obtained by integrating the away-side low-multiplicity $\Delta\varphi$ correlation function in the low-multiplicity sample over $\pi/2 < \Delta\varphi < 3\pi/2$.
As PYTHIA does not include any flow contributions in its model, $Y^{\rm{away}}$ can be directly measured from the $\Delta\varphi$ correlation functions.

\begin{figure}[h!]
	\centering
	\hspace{-3em}\includegraphics[width=0.6\textwidth]{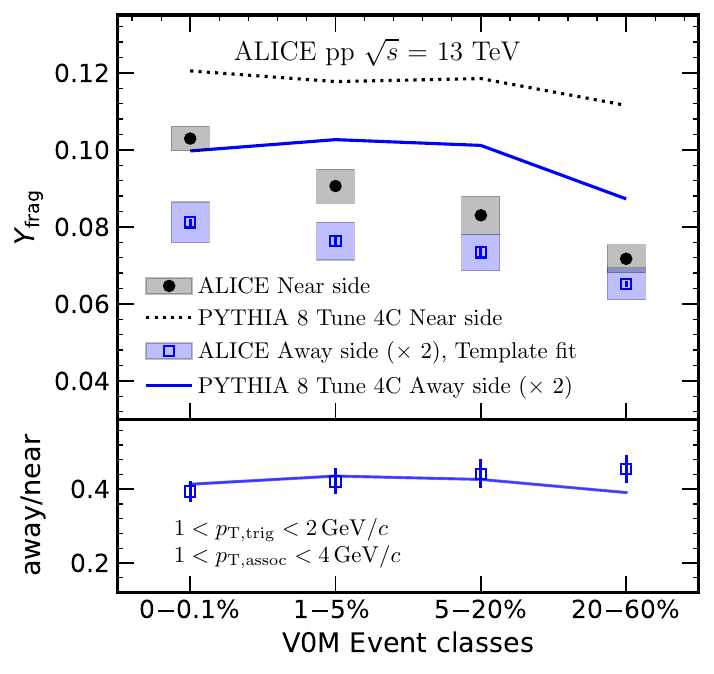} 
	\caption{The $Y_{\rm{frag}}$ for the near- and away-side as a function of multiplicity percentiles with both ALICE and PYTHIA data. Systematic uncertainties are represented by the boxes. The bottom panel presents ratios of the yields. Here the reported uncertainty is obtained by adding statistical error and systematic uncertainty in quadrature. For PYTHIA, the statistical uncertainty is smaller than the thickness of the lines.}
	\label{fig:Ymult}
\end{figure}

Figure~\ref{fig:Ymult} presents the $Y^{\mathrm{near}}_{\rm{frag}}$ and $Y^{\mathrm{away}}_{\rm{frag}}$, for both ALICE data and PYTHIA 8 Tune 4C~\cite{Skands:2014pea}, as a function of the V0M multiplicity intervals in pp collisions at $\sqrt{s}=13$~TeV. The transverse momentum range for trigger particles is $1<p_\mathrm{T,trig}<2$ GeV/$c$ and for associated particles $1<p_\mathrm{T,assoc}<4$ GeV/$c$.
The near- to away-side ratio for ALICE and PYTHIA data is shown in the bottom panel. While PYTHIA overestimates both near-side and away-side yields measured by ALICE, the corresponding ratio is consistent with the ALICE data in the all considered V0M multiplicity intervals. The value of this ratio can be explained by the pair acceptance effect caused by the limited ALICE $\eta$ acceptance~\cite{PHENIX:2006gto}. The observed agreement implies that the enhanced jet fragmentation yields in the away-side in high-multiplicity events with respect to low-multiplicity events~\cite{ALICE:2013tla,ALICE:2014mas} are taken into account by the low-multiplicity template method. In summary, the difference between the near-side and away-side jet fragmentation yields in PYTHIA is solely caused by the jet acceptance effects which affect the two-particle correlation functions. The corresponding ratio in data, where the away-side jet fragmentation yields are measured with the low-multiplicity template, agrees well with PYTHIA as well as with the expectation in Ref.~\cite{PHENIX:2006gto}.

The flow coefficients, $v_{n}$, of the trigger particles, can be extracted from the template fit with the use of the observed factorization of $v_{n,n}$ coefficients to single harmonics~\cite{ATLAS:2015hzw,ATLAS:2016yzd} by using
\begin{eqnarray}
v_{n}(p_{\rm{T,trig}}) = v_{n,n}(p_{\rm{T,trig}}, p_{\rm{T,assoc}}) / \sqrt{ v_{n,n}(p_{\rm{T,assoc}},p_{\rm{T,assoc}})},
\end{eqnarray}
where $v_{n,n}(p_{\rm{T,assoc}}, p_{\rm{T,assoc}})$ denote $v_{n,n}$ coefficients extracted using trigger and associated particles with $p_{\rm{T}}$ in the range 1--4~GeV/$c$. In the following sections, unless explicitly stated otherwise, $v_n$ will refer to $v_n(p_{\rm{T,trig}})$.
Different event scale selections were investigated by selecting events that include a hard jet or a high-$\pt$ leading particle at midrapidity (i.e., the particle with the highest reconstructed $\pt$ inside the acceptance region in an event).
This event scale was set by requiring a minimum $\pt$ of the leading track ($\ptlead$) or the reconstructed jet ($\ptjet$) at midrapidity~\cite{ALICE:2021nir}. The leading particle track was required to be within $|\eta|<0.9$ and $0<\varphi<2\pi$, and the jets were reconstructed with the anti-$k_{\rm{T}}$ algorithm~\cite{Cacciari:2008gp,Cacciari:2011ma}, with $R=0.4$ using charged particles only. Jet constituents were combined using the boost-invariant $\pt$ recombination scheme. The jets are selected in the full azimuth ($0<\varphi<2\pi$) and their pseudorapidity is constrained to $|\eta_\mathrm{jet}|<0.4$. The $\pt$ of jets $\ptjet$ is corrected for the underlying event density that is measured using the $k_{\rm{T}}$ algorithm with $R=$~0.2 following the procedure described in Ref.~\cite{Acharya:2018eat}.


\section{Systematic uncertainties}
\label{sec:uncertainties}

\begin{table}[h!]
\caption{The relative systematic uncertainties of $Y^{\rm{near}}$, $Y^{\rm{away,LM}}$, $F$, $v_{2}$, and $v_{3}$. The quantities $Y^{\rm{near}}$, $Y^{\rm{away,LM}}$, $F$ are only measured in pp collisions, whereas $v_{2}$ and $v_{3}$ are measured in both pp and p--Pb collisions. The quoted ranges correspond to minimum and maximum uncertainties. Those uncertainties that are considered to be negligible are marked ``negl.". The systematic variations which are not relevant for the measurement are denoted as ``N.A".}
\centering
\label{tab:syst}
\resizebox{\textwidth}{!} {
\begin{tabular}{c|ccc|cccc}
\hline 
\multirow{3}{*}{Sources}  & \multicolumn{7}{c}{Systematic uncertainty (\%)} \\ \cline{2-8} 
& \multicolumn{1}{c}{$Y^{\rm{near}}$} & \multicolumn{1}{c}{$Y^{\rm{away,LM}}$} & \multicolumn{1}{c|}{$F$} & \multicolumn{2}{c}{$v_{2}$} & \multicolumn{2}{c}{$v_{3}$}  \\  \cline{2-8}
&pp &pp &pp & pp & p--Pb & pp & p--Pb  \\ \cline{1-8} 
Primary vertex       & $\pm$0.2--0.5 & $\pm$0.1      & $\pm$1.0--2.5 & $\pm$0.2--1.8 & $\pm$0.8 & $\pm$1.4 & $\pm$3.9 \\ 
Pileup rejection     & $\pm$0.1--0.5 & $\pm$0.2      & $\pm$0.4--1.5 & negl.         & $\pm$0.6 & negl. & $\pm$1.4 \\ 
Tracking		     & $\pm$1.0--3.0 & $\pm$2.0      & $\pm$0.6--2.4 & $\pm$0.2--3.0 & negl. & $\pm$5.0--6.9 & negl. \\ 
Event mixing	     & $\pm$0.2--0.7 & $\pm$0.2--0.5 & $\pm$0.0--3.3 & $\pm$0.3--4.6 & $\pm$0.8 & $\pm$2.8--3.1 & $\pm$0.8 \\ 
Low-mult. definition & N.A.          & $\pm$0.5--3.5 & $\pm$0.7--6.0 & negl.         & $\pm$1.9 & negl. & $\pm$9.2\\ 
ITS--TPC matching 	 & $\pm$2.0--3.0 & $\pm$2.0--3.0 & N.A.          & N.A.          & N.A. & N.A. & N.A\\ 
Efficiency correction& $\pm$1.0--4.4 & $\pm$1.0--4.4 & N.A.          & N.A.          & N.A. & N.A. & N.A\\ 
$\Delta\eta$ gap range   	 & N.A.          & N.A.          & $\pm$0.1--3.2 & $\pm$1.0--5.0 & $\pm$0.4 & negl. & negl. \\ 
Jet shape modification   	 & N.A.          & N.A.          & N.A &  $\pm$1.0& $\pm$1.0 & $\pm$3.0 & $\pm$8.0 \\ 
\hline 
Total (in quadrature)& $\pm$2.5--6.1 & $\pm$5.0--5.5 & $\pm$1.8--7.1 & $\pm$1.3--5.8 & $\pm$2.5 & $\pm$6.8--8.0 & $\pm$12.8 \\ 
\hline 
\end{tabular}
}
\end{table}

Systematic uncertainties are estimated by varying the analysis selection criteria and corrections. Independent systematic checks are performed and the differences between results obtained from each variation and the default selection are considered as the systematic uncertainty for each source~\cite{Barlow:2002yb}. The total systematic uncertainty is obtained by adding the contributions from different sources in quadrature. A summary of all systematic uncertainties is provided in Table~\ref{tab:syst}. 

The uncertainty attributed to the chosen primary vertex range is estimated by varying the selected range from $|z_\mathrm{vtx}|<$~8~cm to $|z_\mathrm{vtx}|<$~6~cm. The variation of the range allows testing detector acceptance effects on the measurement.  

Another source of systematic uncertainty is related to pileup rejection. The rejection of pileup events is adjusted by modifying the number of track contributors required for the reconstruction of pileup event vertices, changing it from the default value of 3 to 5.

The systematic uncertainty due to the choice of track-selection criteria is estimated by employing alternative track-selection criteria, which single out so-called “global track”, which are described in Ref.~\cite{ALICE:2021ptz}. A global track is required to have two hits in the ITS (at least one in the SPD) and at least 70 clusters in the TPC. Due to inefficiencies in certain parts of the SPD, the azimuthal distribution of global tracks is not uniform. This can be corrected by using corresponding mixed events and accounting for the corresponding tracking efficiency.

An additional systematic uncertainty from the event-mixing is estimated by varying the interval of the primary vertex range, where events are mixed. The default size of the primary vertex bins of mixed events is decreased from 2 cm to 1 cm. 

The systematic uncertainty associated with the low-multiplicity definition is estimated by changing the range of the low-V0M-multiplicity interval. There is no universal definition for the low-multiplicity interval. The default range for the low-multiplicity interval in the present article is 60--100\%, and it is changed to 70--100\% to estimate the related systematic uncertainty. 
Note that for the measurement of $Y^{\rm{near}}$, the low-multiplicity-interval definition is irrelevant. 

The systematic uncertainty resulting from matching a track reconstructed by the TPC and the corresponding signal in the ITS is estimated by evaluating the fraction of mismatches in real data using simulations. Primary tracks have a higher matching efficiency than secondary tracks produced far from the interaction point or in interactions with detector material. To address the effect of different fractions of primary and secondary tracks in data and simulations on the matching efficiency, particle abundances in the simulation are reweighted to reflect real data. This resulted in modified matching efficiency.

The systematic uncertainty arising from the efficiency correction is estimated by performing a closure test, where two correlation functions are compared. The first correlation function is constructed using true information from the Monte Carlo samples described in Sec.~\ref{sec:experiment}. This provides a baseline for the expected correlation function. The second correlation function is constructed using reconstructed tracks corrected for tracking efficiency. By comparing the two correlation functions, it is possible to estimate the magnitude of the uncertainty introduced by the correction. 

Due to the limited $\eta$ acceptance of the TPC, non-flow contributions, mainly originating from fragmentation of jets, affect the flow measurement. As the shape of short-range correlations, mostly attributed to jets, is getting broader with decreasing $\pt$, the systematic uncertainty from $\eta$ acceptance significantly depends on $p_{\rm{T}}$. To estimate the related uncertainty, the long range $\Delta\varphi$ correlations are measured for an extended $\Delta\eta$ gap, the default size of $\Delta\eta$ gap 1.6 is increased to 1.7.

Finally, it is worth considering the possible impact of the multiplicity dependence of the jet-shape modifications discussed in Sec.~\ref{sec:ana}. This is studied by examining the shape modification of the jet-peak distribution in the near-side region as a function of $\Delta\eta$ and multiplicity. The observed change in the width is used to estimate the possible effect on the long-range, per-trigger-yield distribution as a function of $\Delta\varphi$. The effect on $v_2$ is found to be less than 1$\%$ in pp and p--Pb collisions for the kinematic ranges analyzed. The effect on $v_3$ is found to be less than 8\% in p--Pb collisions. These values are included in the total systematic uncertainty. However, it is important to note that other analyses with different kinematic ranges should also perform a similar check to assess the systematic uncertainty associated with this effect. It is possible that this effect may not always be as small as in our analysis.

\section {Results}
\label{sec:results}
\subsection{Transverse-momentum and multiplicity dependence of anisotropic flow}

\begin{figure}[!b]
	\centering
	\hspace{-4.5em}\includegraphics[width=0.92\textwidth]{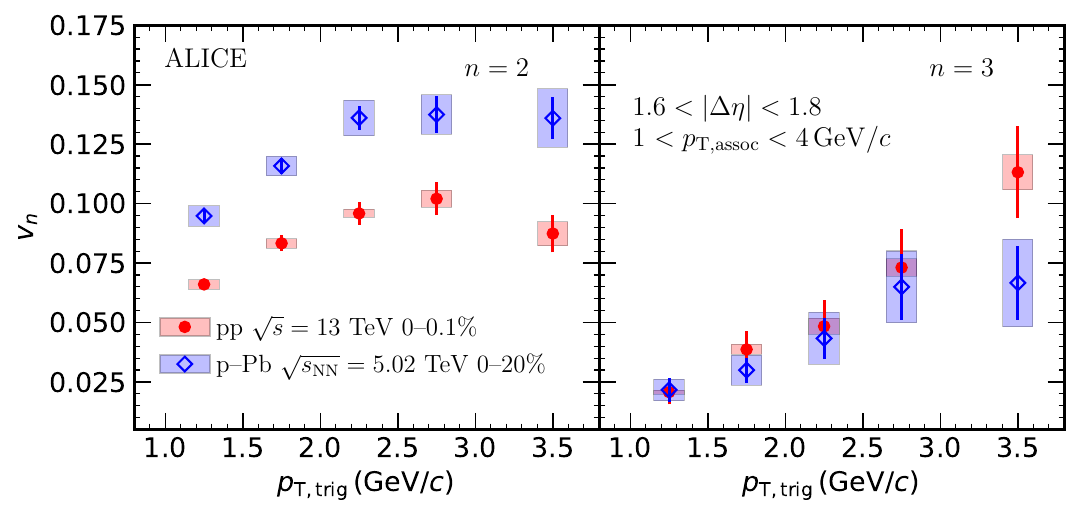} 
	\caption{The magnitude of $v_2$ (left) and $v_3$ (right) as a function of $p_\mathrm{T}$ for the 0--0.1\% multiplicity interval in pp collisions at $\sqrt{s}=13$~TeV and 0--20\% in p--Pb collisions at $\sqrt{s_\mathrm{NN}} = 5.02$~TeV. The boxes around the data points represent the estimated systematic uncertainty and the error bars correspond to statistical errors.}
	\label{fig:vn}
\end{figure}
Figure~\ref{fig:vn} illustrates the extracted values of $v_2$ and $v_3$ as a function of $p_{\mathrm{T,trig}}$ as obtained from Eq.~\eqref{eq:tmpfit}. The results correspond to the high-multiplicity pp collisions at $\sqrt{s}=13$~TeV and p--Pb collisions at $\sqrt{s_\mathrm{NN}}=$5.02~TeV. Both sets of results demonstrate an increasing trend in the magnitudes of $v_n$ with rising $p_{\mathrm{T,trig}}$. The $v_2$ data points reach a maximum between 2.5 and 3.0 GeV/$c$, similarly to what has been observed in Pb--Pb collisions~\cite{Abelev:2012di, ALICE:2018yph}.
The magnitudes of $v_2$ in p--Pb collisions are higher than those in pp collisions, which might be related to the larger p--Pb system size together with a likely longer lifetime of the hypothetically created medium.
However, the magnitudes of $v_3$ are similar in both systems, indicating that $v_3$ is less sensitive to the size of the systems.
These results are comparable to those obtained by ATLAS in different multiplicity classes, where the same method was used to extract the flow coefficients~\cite{ATLAS:2016yzd}. Even though the $\Delta\eta$ and $p_{\mathrm{T,assoc}}$ ranges used by ATLAS are wider, $2.0<|\Delta\eta|<5.0$ and $0.5<p_{\mathrm{T,assoc}}<5\,\mathrm{GeV}/c$, respectively, the results are consistent within uncertainties.

\begin{figure}[!t]
	\centering
\includegraphics[width=1.0\textwidth]{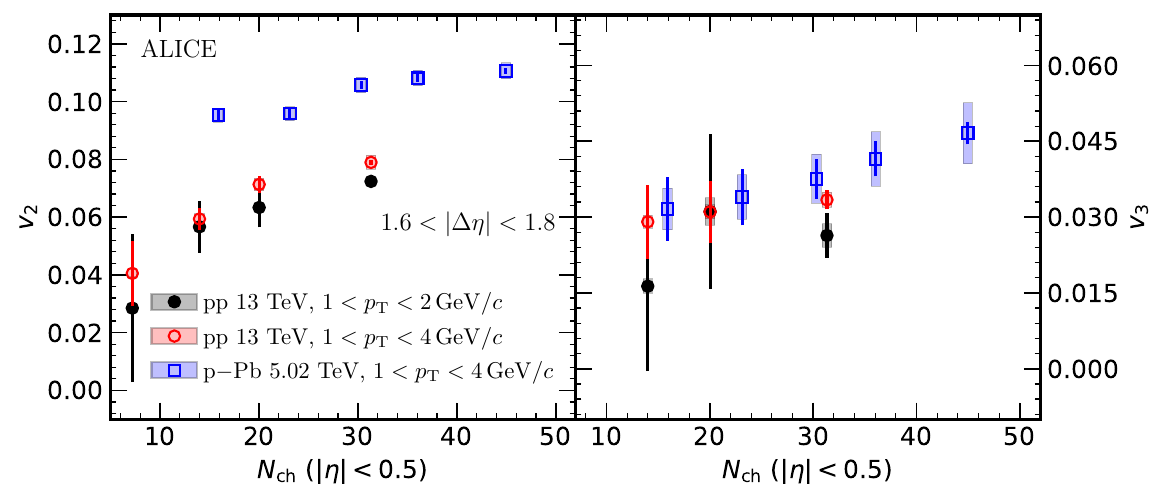} 
	\caption{The magnitudes of $v_2$ (left
    panel) and $v_3$ (right panel) 
 for two different collision systems, pp and p--Pb as a function of charged-particle multiplicity at midrapidity. For pp collisions, two different $p_\mathrm{T}$ intervals are shown, $1.0<p_\mathrm{T}<2.0$ GeV/$c$ and $1.0<p_\mathrm{T}<4.0$ GeV/$c$. The boxes around the data points represent the estimated systematic uncertainties and the error bars corresponds to statistical errors.} 
	\label{fig:v2mult}
\end{figure}

Figure~\ref{fig:v2mult} presents the magnitudes of $v_2$ and $v_3$, as a function of charged-particle multiplicity at midrapidity for both pp collisions at $\sqrt{s}=13$ TeV and p--Pb collisions at $\sqrt{s_\mathrm{NN}}=5.02$ TeV. As in Fig.~\ref{fig:vn}, the $\Delta\eta$ gap is $1.6<|\Delta\eta|<1.8$ and $v_2$ is measured in $1<p_{\mathrm{T}}<4\,\mathrm{GeV}/c$ for both collision systems. Additionally, the corresponding results for pp collisions at $\sqrt{s}=13$ TeV with $1<p_{\mathrm{T}}<2$ GeV/$c$ are presented. First, it is observed that the magnitude of $v_n$ increases with increasing multiplicity for both collision systems and $p_\mathrm{T}$-ranges. Second, $v_2$ in p--Pb is higher than in pp collisions in the measured multiplicity range. These two observations are compatible with previous results from Refs.~\cite{ATLAS:2015hzw,ATLAS:2016yzd, Khachatryan:2015lva}. There is no significant difference between the two collision systems when considering the $v_3$ dependence on multiplicity as shown on the right-hand-side panel of Fig.~\ref{fig:v2mult}. The $v_3$ measurements exhibit a comparable subtle dependence on multiplicity, with higher values observed in collisions with greater particle multiplicities.
For the two different $p_\mathrm{T}$ intervals presented for the pp collisions, the $v_2$ in $1.0<p_\mathrm{T}<4.0$ GeV/$c$ shows a hint of larger magnitude than the $v_2$ in $1.0<p_\mathrm{T}<$~2.0~GeV/$c$. The difference in magnitude is significant only in the highest multiplicity point. 
This agrees with what is observed in Fig.~\ref{fig:vn}, where the $v_2$ magnitude has its largest value in $2.5<p_\mathrm{T}<$~3.0~GeV/$c$. It is found that for the considered $p_\mathrm{T}$ selections, the observed multiplicity dependencies differ only marginally. 
It is worth noting that the results presented from pp and p--Pb collisions were obtained from two different beam energies. In Ref.~\cite{ATLAS:2016yzd}, it was found that the magnitudes of $v_2$ and $v_3$ in pp collisions between 13 and 5.02 TeV show no significant variation with center-of-mass energy.

\subsection{Event-scale dependence of the flow coefficients}
\begin{figure}[h!]
	\centering
	\hspace{-3em}\includegraphics[width=0.8\textwidth]{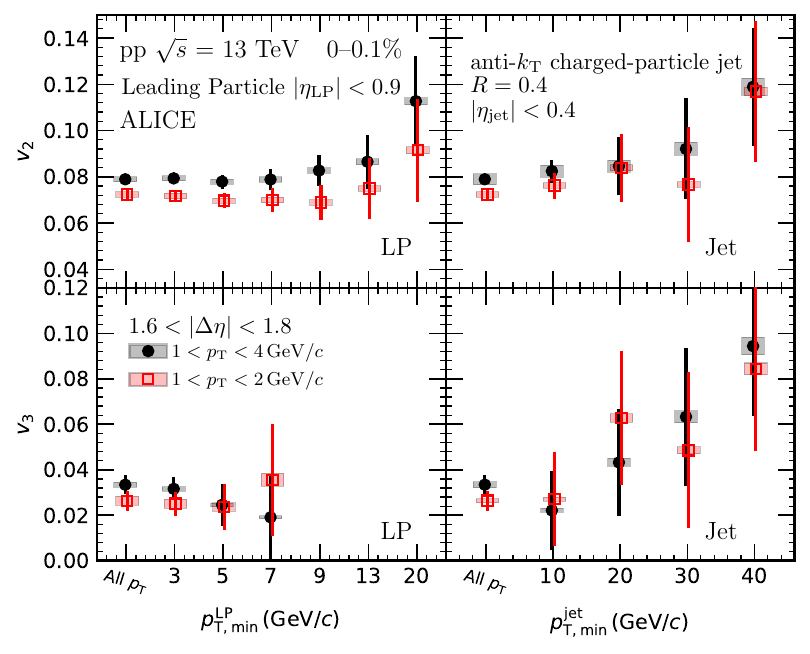}
	\caption{The magnitudes of $v_2$ (top) and $v_3$ (bottom) as a function of $\it{p}^{\rm{LP}}_{\rm{T,min}}$ (left) and $\it{p}^{\rm{jet}}_{\rm{T,min}}$ (right) for the high-multiplicity in pp collisions at $\sqrt{s}=13$ TeV. The measured $p_{\mathrm{T}}$ intervals are $1<p_{\mathrm{T}}<2\,\mathrm{GeV}/c$ (in red) and $1<p_{\mathrm{T}}<4\,\mathrm{GeV}/c$ (in black). The statistical errors and systematic uncertainties are shown as vertical bars and boxes, respectively.}
	\label{fig:LPjet23}
\end{figure}    

Figure~\ref{fig:LPjet23} presents the extracted magnitude of $v_2$ and $v_3$ as a function of the minimum $p_\mathrm{T}$ of the leading particle $\it{p}^{\rm{LP}}_{\rm{T,min}}$ and that of the jet ($\it{p}^{\rm{jet}}_{\rm{T,min}}$) as introduced in Sec.~\ref{sec:intro}. 
The results are presented for the 0–0.1\% multiplicity class of pp collisions at $\sqrt{s}= 13$ TeV and for the two different $p_\mathrm{T}$-ranges.
To reduce the impact of the detector edge effects on the jet measurements, the jet axes are required to have a pseudorapidity $|\eta_\mathrm{jet}|<0.4$, following a similar selection as in Refs.~\cite{CDF:2001onq, ATLAS:2014riz, CMS:2015jdl}. The $v_2$ and $v_3$ values for both $p_\mathrm{T}$ ranges do not show any dependence on event-scale selection within the uncertainties. This finding is consistent with the results of the ridge yields~\cite{ALICE:2021nir} and $v_{2}$ measurements with a tagged $Z$ boson from the ATLAS collaboration~\cite{Aaboud:2019mcw}. These results suggest that the presence of a hard-scattering process does not significantly change the long-range correlation involving soft particles.
However, the presented measurements are only limited to the low $\it{p}^{\rm{jet}}_{\rm{T}}$. Future measurements with multi-jet events at midrapidity with higher $Q^2$ reach can shed more light on the expected impact parameter dependence~\cite{Sjostrand:1986ep,Frankfurt:2003td,Frankfurt:2010ea}.

\subsection{Comparisons with models}
\label{sec:theory}

\begin{figure}[h!]
	\centering
	\includegraphics[width=1.0\textwidth]{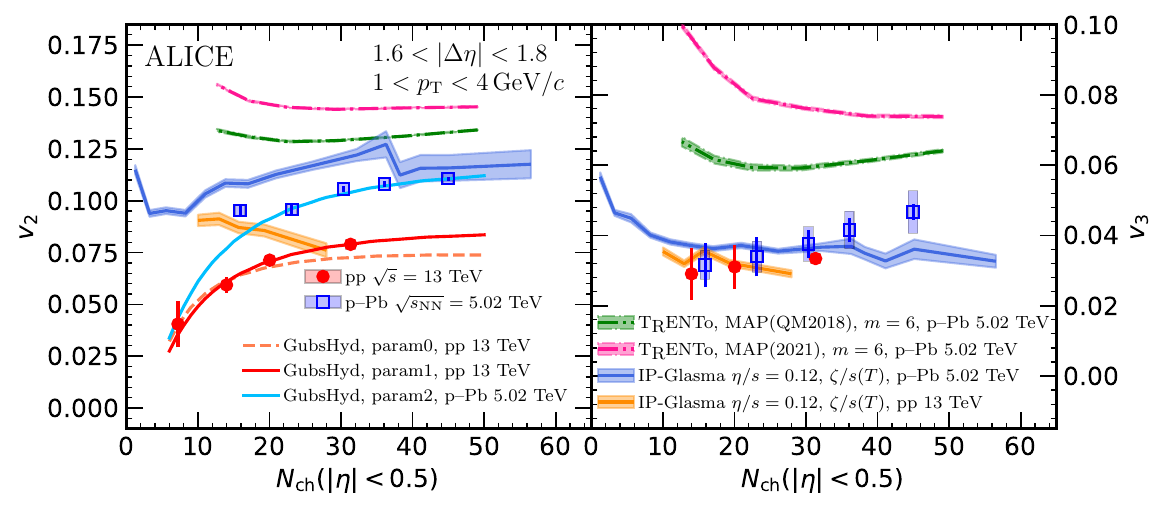} 
	\caption{The measured and calculated evolution of $v_2$ (left) and $v_3$ (right) in pp and p--Pb collisions as a function of charged-particle multiplicity at midrapidity.
The blue and red markers represent the measured p--Pb and pp data, respectively. The calculations provided by hydrodynamical models~\cite{Parkkila:2021yha,Bernhard:2016tnd,Schenke:2020mbo,Taghavi:2019mqz} are presented with colored lines. The corresponding bands mark their statistical uncertainty. For GubsHyd calculations, the statistical uncertainty is smaller than the line thickness.} 
	\label{fig:vnmult_model}
\end{figure}

In this section, the results are compared to various model calculations.
The results from p--Pb collisions are compared with hydrodynamic calculations using the parameterization from an improved global
Bayesian analysis. The analysis involves new sophisticated collective flow observables as obtained from two different beam energies in Pb--Pb collisions~\cite{Parkkila:2021yha}, constraining the initial conditions and transport properties of the QGP. This hydrodynamic model, {T\raisebox{-.5ex}{R}ENTo}+iEBE-VISHNU, consists of the {T\raisebox{-.5ex}{R}ENTo} model~\cite{Moreland:2014oya} to simulate the initial condition, which is connected with a free streaming to a 2+1 dimensional causal hydrodynamic model VISH2+1~\cite{Shen:2014vra}. The evolution is continued after hadronization with a hadronic cascade model (UrQMD)~\cite{Bass:1998ca,Bleicher:1999xi}. 
A model calculation is performed using the best-fit parameterization for transport coefficients selected based on maximum a posteriori (MAP) for Pb--Pb collisions at $\sqrt{s_{\text{NN}}}=5.02$~TeV. Two different MAP values are used for the calculations. They are based on Ref.~\cite{Parkkila:2021yha} and Ref.~\cite{Bernhard:2016tnd} and in Fig. 7 they
are labeled MAP(2021) and MAP(QM2018), respectively.
 The parameterization for the initial conditions, which include a sub-nucleon structure with six constituent partons per nucleon ($m=6$), is taken from a model calibration with additional p--Pb data~\cite{Moreland:2018gsh}. All kinematic selections, such as the transverse momentum and pseudorapidity intervals, are matched to the data reported in this article. The flow coefficients in the hydrodynamic calculation are extracted with the two-particle cumulant method, as the {T\raisebox{-.5ex}{R}ENTo}+iEBE-VISHNU does not contain any non-flow.

Figure~\ref{fig:vnmult_model} shows that {T\raisebox{-.5ex}{R}ENTo}+iEBE-VISHNU overestimates both $v_2$ and $v_3$. In the studied range, the $v_2$ and $v_3$ data increase with multiplicity. However, {T\raisebox{-.5ex}{R}ENTo}+iEBE-VISHNU predicts the opposite trend, which is similar to what is found in large collision systems~\cite{Acharya:2020taj}. The large discrepancies in the prediction might be alleviated by inclusion of the newly measured p--Pb constraints in a future Bayesian parameter estimation as well as by improvements of the initial condition model for small-system collisions.

The results are also compared with IP-Glasma+MUSIC+UrQMD hydrodynamic calculations~\cite{Schenke:2020mbo}. This model uses IP-Glasma initial conditions~\cite{Schenke:2012wb} including sub-nucleonic fluctuations with three hot spots per nucleon. The hydrodynamic evolution is performed by MUSIC~\cite{Schenke:2010rr} and coupled with UrQMD~\cite{Bass:1998ca,Bleicher:1999xi}, which performs hadronic cascade. 
The model calculations are performed assuming constant $\eta/s=0.12$ and a temperature dependent $\zeta/s(T)$~\cite{Rose:2020lfc}. 
This model describes well the multiplicity dependence of $v_2$ in p--Pb collisions and the magnitude at the highest multiplicity within the statistical uncertainties of the model but overestimates the data for the lower multiplicity classes. As for pp collisions, the calculations clearly miss both the observed magnitude except for $N_\mathrm{ch}>25$ as well as the trend of the multiplicity dependence. The model shows that $v_2$ decreases with increasing multiplicity, while the experimental result shows the opposite.
For $v_3$, the model accurately describes the magnitudes and multiplicity dependence across the measured multiplicity ranges. The magnitudes of $v_3$ are slightly smaller in pp collisions than in p--Pb collisions according to the calculations, which agrees with the data within the uncertainties. The level of agreement between data and the IP Glasma model calculations is found to be similar to the results reported in Ref.~\cite{Acharya:2019vdf}.

Finally, the results are compared with the GubsHyd model, a semi-analytical model based on the analytical Gubser solution to hydrodynamic equations~\cite{Gubser:2010ze,Gubser:2010ui}, known as Gubser flow. In Gubser flow, the initial state of conformal matter is linearly perturbed by an initial elliptic shape. The model is employed to shed light on the possible sources of the observed discrepancy between more realistic models mentioned above and the measurements in pp collisions~\cite{Taghavi:2019mqz}. Instead of modeling the initial entropy density in this model, as it is typically done in T$_{\text{R}}$ENTo or IP-Glasma, the initial state fluctuation is modeled directly. It assumes that proton ellipticity $\epsilon_{2}$ and RMS radius $r_{\text{rms}}$ fluctuate independently. These fluctuations are described by Gaussian probability distributions, which have widths $\sigma_{\epsilon}$
 and $\sigma_{r}$, respectively. The multiplicity dependence of the $v_2\{2\}$ of two-particle correlation functions depends on $\sigma_{r}$ and  $\chi\sigma_{\epsilon}$, where the coefficient $\chi$ encapsulates a correction for idealizations used in GubsHyd, including the absence of dissipation effects. The values of $\sigma_{r}$ and  $\chi\sigma_{\epsilon}$ were obtained by comparing the model with data. Since no non-flow effect is considered in the calculation, $v_2\{2\}$ is comparable with the flow measurements in the present study. The calculations for two sets of parameters are compared
to data in Fig.~\ref{fig:vnmult_model}. The “param0” parameterization is based on the prediction proposed in Ref.~\cite{Taghavi:2019mqz} that $\chi \sigma_{\epsilon} =$ 0.097 and $\sigma_{r} =$ 0.4 fm. The other parameterizations “param1” and “param2” employ different $\chi \sigma_{\epsilon}$  and $\sigma_{r}$ values. The model captures the multiplicity dependence of $v_2$ well.

In summary, the measured $v_{2}$ value decrease with decreasing multiplicity in both pp and p--Pb collisions. This trend is also predicted by GubsHyd model calculations (Refs.~\cite{Taghavi:2019mqz} and~\cite{Weller:2017tsr}). Interestingly, the opposite trend is observed for the IP-GLASMA+MUSIC+UrQMD hydrodynamic calculations of $v_2$, where the value decreases with increasing charged-particle multiplicity~\cite{Schenke:2020mbo}.
Approaching a lower bound for the size of a hydrodynamized
system as predicted in Ref.~\cite{Taghavi:2019mqz}, 
the decreasing trend of $v_2$, obtained by lowering the charged-particle multiplicity, changes and turns out to raise again after the observed minimum. However, this change in multiplicity dependence of $v_2$ at low multiplicities is still challenging to test with the current experimental uncertainties. For $v_3$, the IP-Glasma+MUSIC+UrQMD hydrodynamic calculations~\cite{Schenke:2020mbo} are the only ones that accurately describe its magnitude and multiplicity dependence across the measured ranges. The calculations predict that the magnitudes of $v_3$ are slightly smaller in pp collisions than in p--Pb collisions, within the measured multiplicity ranges. The discrepancies between the predictions and the data can be further studied by including these measurements in a future Bayesian parameter estimation, as well as by improving the initial-condition model for the small-system collisions.
 
\section{Conclusions}
\label{sec:summary}
Long-range angular correlations for pairs of charged particles are studied in pp collisions at ${\sqrt{{\textit s}}}=13$~TeV and p--Pb collisions at $\sqrt{s_\mathrm{NN}} = 5.02$~TeV. Flow coefficients are extracted from long-range correlations ($1.6 <|\Delta\eta| < 1.8$) for a broad range of charged-particle-multiplicity classes using the template method, which allows one to subtract the enhanced away-side jet fragmentation yields in high-multiplicity events with respect to low-multiplicity events. The method that was used to measure the flow coefficients within the considered kinematic ranges has been verified to be stable.
The systematic uncertainties on $v_2$ and $v_3$ measurements, which reflect the possible differences in the away-side jet peak shapes in high- and low-multiplicity events, were found to be 1\% and 3–8\%, respectively.However, it is important that these systematic uncertainties are reevaluated, when analyzing different kinematic ranges, as the effect may not always be negligible.
The measured $p_\mathrm{T}$ dependence of $v_2$ and $v_3$ is consistent with the measurements by ATLAS and shows that both $v_2$ and $v_3$ increase with $p_\mathrm{T}$ and reach their maximum at $2.5<p_\mathrm{T}<3.0\,~\mathrm{GeV}/c$.
The measurement of $v_2$ as a function of charged-particle multiplicity in $|\eta|<0.5$ shows a weak multiplicity dependence both for pp and p--Pb collisions and tends to decrease toward lower multiplicities.
The pp data suggests that the $v_2$ signal may disappear when the measurement is pursued further below $N_{\mathrm{ch}} = 10$. 

The comparisons to viscous hydrodynamic models show that the magnitudes of $v_2$ and their multiplicity dependence are not described by state-of-the-art hydrodynamic calculations, which simulated initial conditions with two initial state models, especially for low-multiplicity p--Pb and pp collisions. As initial state effects tend to be more important at low multiplicity~\cite{Greif:2017bnr,Moreland:2018gsh}, these results may help to constrain the modeling of the initial state.
Furthermore, the events including hard probes such as jets or high-$p_\mathrm{T}$ leading particles do not show any changes both in $v_2$ and $v_3$ within the uncertainties, which implies that the long-range correlation of soft particles is not significantly modified by the presence of the hard-scattering process. Even though it would be interesting to compare these results to the EPOS LHC~\cite{Pierog:2013ria} and PYTHIA8 String Shoving models~\cite{Bierlich:2017vhg,Bierlich:2019ixq} as done in Ref.~\cite{ALICE:2012eyl} for the ridge yields, it is not possible to reliably extract the flow coefficients because these models exhibit a near-side ridge structure in low-multiplicity events, thus making the use of the low-multiplicity template ill defined~\cite{Ji:2023eqn}.

\newenvironment{acknowledgement}{\relax}{\relax}
\begin{acknowledgement}
\section*{Acknowledgements}

The ALICE Collaboration would like to thank all its engineers and technicians for their invaluable contributions to the construction of the experiment and the CERN accelerator teams for the outstanding performance of the LHC complex.
The ALICE Collaboration gratefully acknowledges the resources and support provided by all Grid centres and the Worldwide LHC Computing Grid (WLCG) collaboration.
The ALICE Collaboration acknowledges the following funding agencies for their support in building and running the ALICE detector:
A. I. Alikhanyan National Science Laboratory (Yerevan Physics Institute) Foundation (ANSL), State Committee of Science and World Federation of Scientists (WFS), Armenia;
Austrian Academy of Sciences, Austrian Science Fund (FWF): [M 2467-N36] and Nationalstiftung f\"{u}r Forschung, Technologie und Entwicklung, Austria;
Ministry of Communications and High Technologies, National Nuclear Research Center, Azerbaijan;
Conselho Nacional de Desenvolvimento Cient\'{\i}fico e Tecnol\'{o}gico (CNPq), Financiadora de Estudos e Projetos (Finep), Funda\c{c}\~{a}o de Amparo \`{a} Pesquisa do Estado de S\~{a}o Paulo (FAPESP) and Universidade Federal do Rio Grande do Sul (UFRGS), Brazil;
Bulgarian Ministry of Education and Science, within the National Roadmap for Research Infrastructures 2020-2027 (object CERN), Bulgaria;
Ministry of Education of China (MOEC) , Ministry of Science \& Technology of China (MSTC) and National Natural Science Foundation of China (NSFC), China;
Ministry of Science and Education and Croatian Science Foundation, Croatia;
Centro de Aplicaciones Tecnol\'{o}gicas y Desarrollo Nuclear (CEADEN), Cubaenerg\'{\i}a, Cuba;
Ministry of Education, Youth and Sports of the Czech Republic, Czech Republic;
The Danish Council for Independent Research | Natural Sciences, the VILLUM FONDEN and Danish National Research Foundation (DNRF), Denmark;
Helsinki Institute of Physics (HIP), Finland;
Commissariat \`{a} l'Energie Atomique (CEA) and Institut National de Physique Nucl\'{e}aire et de Physique des Particules (IN2P3) and Centre National de la Recherche Scientifique (CNRS), France;
Bundesministerium f\"{u}r Bildung und Forschung (BMBF) and GSI Helmholtzzentrum f\"{u}r Schwerionenforschung GmbH, Germany;
General Secretariat for Research and Technology, Ministry of Education, Research and Religions, Greece;
National Research, Development and Innovation Office, Hungary;
Department of Atomic Energy Government of India (DAE), Department of Science and Technology, Government of India (DST), University Grants Commission, Government of India (UGC) and Council of Scientific and Industrial Research (CSIR), India;
National Research and Innovation Agency - BRIN, Indonesia;
Istituto Nazionale di Fisica Nucleare (INFN), Italy;
Japanese Ministry of Education, Culture, Sports, Science and Technology (MEXT) and Japan Society for the Promotion of Science (JSPS) KAKENHI, Japan;
Consejo Nacional de Ciencia (CONACYT) y Tecnolog\'{i}a, through Fondo de Cooperaci\'{o}n Internacional en Ciencia y Tecnolog\'{i}a (FONCICYT) and Direcci\'{o}n General de Asuntos del Personal Academico (DGAPA), Mexico;
Nederlandse Organisatie voor Wetenschappelijk Onderzoek (NWO), Netherlands;
The Research Council of Norway, Norway;
Commission on Science and Technology for Sustainable Development in the South (COMSATS), Pakistan;
Pontificia Universidad Cat\'{o}lica del Per\'{u}, Peru;
Ministry of Education and Science, National Science Centre and WUT ID-UB, Poland;
Korea Institute of Science and Technology Information and National Research Foundation of Korea (NRF), Republic of Korea;
Ministry of Education and Scientific Research, Institute of Atomic Physics, Ministry of Research and Innovation and Institute of Atomic Physics and University Politehnica of Bucharest, Romania;
Ministry of Education, Science, Research and Sport of the Slovak Republic, Slovakia;
National Research Foundation of South Africa, South Africa;
Swedish Research Council (VR) and Knut \& Alice Wallenberg Foundation (KAW), Sweden;
European Organization for Nuclear Research, Switzerland;
Suranaree University of Technology (SUT), National Science and Technology Development Agency (NSTDA) and National Science, Research and Innovation Fund (NSRF via PMU-B B05F650021), Thailand;
Turkish Energy, Nuclear and Mineral Research Agency (TENMAK), Turkey;
National Academy of  Sciences of Ukraine, Ukraine;
Science and Technology Facilities Council (STFC), United Kingdom;
National Science Foundation of the United States of America (NSF) and United States Department of Energy, Office of Nuclear Physics (DOE NP), United States of America.
In addition, individual groups or members have received support from:
European Research Council, Strong 2020 - Horizon 2020 (grant nos. 950692, 824093), European Union;
Academy of Finland (Center of Excellence in Quark Matter) (grant nos. 346327, 346328), Finland.
\end{acknowledgement}

\bibliographystyle{utphys}
\bibliography{paper.bib}



\newpage
\appendix

\section{The ALICE Collaboration}
\label{app:collab}
\begin{flushleft} 
\small

S.~Acharya\,\orcidlink{0000-0002-9213-5329}\,$^{\rm 128}$, 
D.~Adamov\'{a}\,\orcidlink{0000-0002-0504-7428}\,$^{\rm 87}$, 
G.~Aglieri Rinella\,\orcidlink{0000-0002-9611-3696}\,$^{\rm 33}$, 
M.~Agnello\,\orcidlink{0000-0002-0760-5075}\,$^{\rm 30}$, 
N.~Agrawal\,\orcidlink{0000-0003-0348-9836}\,$^{\rm 52}$, 
Z.~Ahammed\,\orcidlink{0000-0001-5241-7412}\,$^{\rm 136}$, 
S.~Ahmad\,\orcidlink{0000-0003-0497-5705}\,$^{\rm 16}$, 
S.U.~Ahn\,\orcidlink{0000-0001-8847-489X}\,$^{\rm 72}$, 
I.~Ahuja\,\orcidlink{0000-0002-4417-1392}\,$^{\rm 38}$, 
A.~Akindinov\,\orcidlink{0000-0002-7388-3022}\,$^{\rm 142}$
% ...

\section*{Affiliation Notes}

$^{\rm I}$ Deceased\\
$^{\rm II}$ Also at: Max-Planck-Institut fur Physik, Munich, Germany\\
$^{\rm III}$ Also at: Italian National Agency for New Technologies, Energy and Sustainable Economic Development (ENEA), Bologna, Italy\\
$^{\rm IV}$ Also at: Dipartimento DET del Politecnico di Torino, Turin, Italy\\
$^{\rm V}$ Also at: Department of Applied Physics, Aligarh Muslim University, Aligarh, India\\
$^{\rm VI}$ Also at: Institute of Theoretical Physics, University of Wroclaw, Poland\\
$^{\rm VII}$ Also at: An institution covered by a cooperation agreement with CERN\\

\section*{Collaboration Institutes}

$^{1}$ A.I. Alikhanyan National Science Laboratory (Yerevan Physics Institute) Foundation, Yerevan, Armenia\\
$^{2}$ AGH University of Krakow, Cracow, Poland\\
$^{3}$ Bogolyubov Institute for Theoretical Physics, National Academy of Sciences of Ukraine, Kiev, Ukraine\\
$^{4}$ Bose Institute, Department of Physics  and Centre for Astroparticle Physics and Space Science (CAPSS), Kolkata, India\\
$^{5}$ California Polytechnic State University, San Luis Obispo, California, United States\\
$^{6}$ Central China Normal University, Wuhan, China\\
$^{7}$ Centro de Aplicaciones Tecnol\'{o}gicas y Desarrollo Nuclear (CEADEN), Havana, Cuba\\
$^{8}$ Centro de Investigaci\'{o}n y de Estudios Avanzados (CINVESTAV), Mexico City and M\'{e}rida, Mexico\\
$^{9}$ Chicago State University, Chicago, Illinois, United States\\
$^{10}$ China Institute of Atomic Energy, Beijing, China\\
$^{11}$ China University of Geosciences, Wuhan, China\\
$^{12}$ Chungbuk National University, Cheongju, Republic of Korea\\
$^{13}$ Comenius University Bratislava, Faculty of Mathematics, Physics and Informatics, Bratislava, Slovak Republic\\
$^{14}$ COMSATS University Islamabad, Islamabad, Pakistan\\
$^{15}$ Creighton University, Omaha, Nebraska, United States\\
$^{16}$ Department of Physics, Aligarh Muslim University, Aligarh, India\\
$^{17}$ Department of Physics, Pusan National University, Pusan, Republic of Korea\\
$^{18}$ Department of Physics, Sejong University, Seoul, Republic of Korea\\
$^{19}$ Department of Physics, University of California, Berkeley, California, United States\\
$^{20}$ Department of Physics, University of Oslo, Oslo, Norway\\
$^{21}$ Department of Physics and Technology, University of Bergen, Bergen, Norway\\
$^{22}$ Dipartimento di Fisica, Universit\`{a} di Pavia, Pavia, Italy\\
$^{23}$ Dipartimento di Fisica dell'Universit\`{a} and Sezione INFN, Cagliari, Italy\\
$^{24}$ Dipartimento di Fisica dell'Universit\`{a} and Sezione INFN, Trieste, Italy\\
$^{25}$ Dipartimento di Fisica dell'Universit\`{a} and Sezione INFN, Turin, Italy\\
$^{26}$ Dipartimento di Fisica e Astronomia dell'Universit\`{a} and Sezione INFN, Bologna, Italy\\
$^{27}$ Dipartimento di Fisica e Astronomia dell'Universit\`{a} and Sezione INFN, Catania, Italy\\
$^{28}$ Dipartimento di Fisica e Astronomia dell'Universit\`{a} and Sezione INFN, Padova, Italy\\
$^{29}$ Dipartimento di Fisica `E.R.~Caianiello' dell'Universit\`{a} and Gruppo Collegato INFN, Salerno, Italy\\
$^{30}$ Dipartimento DISAT del Politecnico and Sezione INFN, Turin, Italy\\
$^{31}$ Dipartimento di Scienze MIFT, Universit\`{a} di Messina, Messina, Italy\\
$^{32}$ Dipartimento Interateneo di Fisica `M.~Merlin' and Sezione INFN, Bari, Italy\\
$^{33}$ European Organization for Nuclear Research (CERN), Geneva, Switzerland\\
$^{34}$ Faculty of Electrical Engineering, Mechanical Engineering and Naval Architecture, University of Split, Split, Croatia\\
$^{35}$ Faculty of Engineering and Science, Western Norway University of Applied Sciences, Bergen, Norway\\
$^{36}$ Faculty of Nuclear Sciences and Physical Engineering, Czech Technical University in Prague, Prague, Czech Republic\\
$^{37}$ Faculty of Physics, Sofia University, Sofia, Bulgaria\\
$^{38}$ Faculty of Science, P.J.~\v{S}af\'{a}rik University, Ko\v{s}ice, Slovak Republic\\
$^{39}$ Frankfurt Institute for Advanced Studies, Johann Wolfgang Goethe-Universit\"{a}t Frankfurt, Frankfurt, Germany\\
$^{40}$ Fudan University, Shanghai, China\\
$^{41}$ Gangneung-Wonju National University, Gangneung, Republic of Korea\\
$^{42}$ Gauhati University, Department of Physics, Guwahati, India\\
$^{43}$ Helmholtz-Institut f\"{u}r Strahlen- und Kernphysik, Rheinische Friedrich-Wilhelms-Universit\"{a}t Bonn, Bonn, Germany\\
$^{44}$ Helsinki Institute of Physics (HIP), Helsinki, Finland\\
$^{45}$ High Energy Physics Group,  Universidad Aut\'{o}noma de Puebla, Puebla, Mexico\\
$^{46}$ Horia Hulubei National Institute of Physics and Nuclear Engineering, Bucharest, Romania\\
$^{47}$ HUN-REN Wigner Research Centre for Physics, Budapest, Hungary\\
$^{48}$ Indian Institute of Technology Bombay (IIT), Mumbai, India\\
$^{49}$ Indian Institute of Technology Indore, Indore, India\\
$^{50}$ INFN, Laboratori Nazionali di Frascati, Frascati, Italy\\
$^{51}$ INFN, Sezione di Bari, Bari, Italy\\
$^{52}$ INFN, Sezione di Bologna, Bologna, Italy\\
$^{53}$ INFN, Sezione di Cagliari, Cagliari, Italy\\
$^{54}$ INFN, Sezione di Catania, Catania, Italy\\
$^{55}$ INFN, Sezione di Padova, Padova, Italy\\
$^{56}$ INFN, Sezione di Pavia, Pavia, Italy\\
$^{57}$ INFN, Sezione di Torino, Turin, Italy\\
$^{58}$ INFN, Sezione di Trieste, Trieste, Italy\\
$^{59}$ Inha University, Incheon, Republic of Korea\\
$^{60}$ Institute for Gravitational and Subatomic Physics (GRASP), Utrecht University/Nikhef, Utrecht, Netherlands\\
$^{61}$ Institute of Experimental Physics, Slovak Academy of Sciences, Ko\v{s}ice, Slovak Republic\\
$^{62}$ Institute of Physics, Homi Bhabha National Institute, Bhubaneswar, India\\
$^{63}$ Institute of Physics of the Czech Academy of Sciences, Prague, Czech Republic\\
$^{64}$ Institute of Space Science (ISS), Bucharest, Romania\\
$^{65}$ Institut f\"{u}r Kernphysik, Johann Wolfgang Goethe-Universit\"{a}t Frankfurt, Frankfurt, Germany\\
$^{66}$ Instituto de Ciencias Nucleares, Universidad Nacional Aut\'{o}noma de M\'{e}xico, Mexico City, Mexico\\
$^{67}$ Instituto de F\'{i}sica, Universidade Federal do Rio Grande do Sul (UFRGS), Porto Alegre, Brazil\\
$^{68}$ Instituto de F\'{\i}sica, Universidad Nacional Aut\'{o}noma de M\'{e}xico, Mexico City, Mexico\\
$^{69}$ iThemba LABS, National Research Foundation, Somerset West, South Africa\\
$^{70}$ Jeonbuk National University, Jeonju, Republic of Korea\\
$^{71}$ Johann-Wolfgang-Goethe Universit\"{a}t Frankfurt Institut f\"{u}r Informatik, Fachbereich Informatik und Mathematik, Frankfurt, Germany\\
$^{72}$ Korea Institute of Science and Technology Information, Daejeon, Republic of Korea\\
$^{73}$ KTO Karatay University, Konya, Turkey\\
$^{74}$ Laboratoire de Physique Subatomique et de Cosmologie, Universit\'{e} Grenoble-Alpes, CNRS-IN2P3, Grenoble, France\\
$^{75}$ Lawrence Berkeley National Laboratory, Berkeley, California, United States\\
$^{76}$ Lund University Department of Physics, Division of Particle Physics, Lund, Sweden\\
$^{77}$ Nagasaki Institute of Applied Science, Nagasaki, Japan\\
$^{78}$ Nara Women{'}s University (NWU), Nara, Japan\\
$^{79}$ National and Kapodistrian University of Athens, School of Science, Department of Physics , Athens, Greece\\
$^{80}$ National Centre for Nuclear Research, Warsaw, Poland\\
$^{81}$ National Institute of Science Education and Research, Homi Bhabha National Institute, Jatni, India\\
$^{82}$ National Nuclear Research Center, Baku, Azerbaijan\\
$^{83}$ National Research and Innovation Agency - BRIN, Jakarta, Indonesia\\
$^{84}$ Niels Bohr Institute, University of Copenhagen, Copenhagen, Denmark\\
$^{85}$ Nikhef, National institute for subatomic physics, Amsterdam, Netherlands\\
$^{86}$ Nuclear Physics Group, STFC Daresbury Laboratory, Daresbury, United Kingdom\\
$^{87}$ Nuclear Physics Institute of the Czech Academy of Sciences, Husinec-\v{R}e\v{z}, Czech Republic\\
$^{88}$ Oak Ridge National Laboratory, Oak Ridge, Tennessee, United States\\
$^{89}$ Ohio State University, Columbus, Ohio, United States\\
$^{90}$ Physics department, Faculty of science, University of Zagreb, Zagreb, Croatia\\
$^{91}$ Physics Department, Panjab University, Chandigarh, India\\
$^{92}$ Physics Department, University of Jammu, Jammu, India\\
$^{93}$ Physics Program and International Institute for Sustainability with Knotted Chiral Meta Matter (SKCM2), Hiroshima University, Hiroshima, Japan\\
$^{94}$ Physikalisches Institut, Eberhard-Karls-Universit\"{a}t T\"{u}bingen, T\"{u}bingen, Germany\\
$^{95}$ Physikalisches Institut, Ruprecht-Karls-Universit\"{a}t Heidelberg, Heidelberg, Germany\\
$^{96}$ Physik Department, Technische Universit\"{a}t M\"{u}nchen, Munich, Germany\\
$^{97}$ Politecnico di Bari and Sezione INFN, Bari, Italy\\
$^{98}$ Research Division and ExtreMe Matter Institute EMMI, GSI Helmholtzzentrum f\"ur Schwerionenforschung GmbH, Darmstadt, Germany\\
$^{99}$ Saga University, Saga, Japan\\
$^{100}$ Saha Institute of Nuclear Physics, Homi Bhabha National Institute, Kolkata, India\\
$^{101}$ School of Physics and Astronomy, University of Birmingham, Birmingham, United Kingdom\\
$^{102}$ Secci\'{o}n F\'{\i}sica, Departamento de Ciencias, Pontificia Universidad Cat\'{o}lica del Per\'{u}, Lima, Peru\\
$^{103}$ Stefan Meyer Institut f\"{u}r Subatomare Physik (SMI), Vienna, Austria\\
$^{104}$ SUBATECH, IMT Atlantique, Nantes Universit\'{e}, CNRS-IN2P3, Nantes, France\\
$^{105}$ Sungkyunkwan University, Suwon City, Republic of Korea\\
$^{106}$ Suranaree University of Technology, Nakhon Ratchasima, Thailand\\
$^{107}$ Technical University of Ko\v{s}ice, Ko\v{s}ice, Slovak Republic\\
$^{108}$ The Henryk Niewodniczanski Institute of Nuclear Physics, Polish Academy of Sciences, Cracow, Poland\\
$^{109}$ The University of Texas at Austin, Austin, Texas, United States\\
$^{110}$ Universidad Aut\'{o}noma de Sinaloa, Culiac\'{a}n, Mexico\\
$^{111}$ Universidade de S\~{a}o Paulo (USP), S\~{a}o Paulo, Brazil\\
$^{112}$ Universidade Estadual de Campinas (UNICAMP), Campinas, Brazil\\
$^{113}$ Universidade Federal do ABC, Santo Andre, Brazil\\
$^{114}$ Universitatea Nationala de Stiinta si Tehnologie Politehnica Bucuresti, Bucharest, Romania\\
$^{115}$ University of Cape Town, Cape Town, South Africa\\
$^{116}$ University of Derby, Derby, United Kingdom\\
$^{117}$ University of Houston, Houston, Texas, United States\\
$^{118}$ University of Jyv\"{a}skyl\"{a}, Jyv\"{a}skyl\"{a}, Finland\\
$^{119}$ University of Kansas, Lawrence, Kansas, United States\\
$^{120}$ University of Liverpool, Liverpool, United Kingdom\\
$^{121}$ University of Science and Technology of China, Hefei, China\\
$^{122}$ University of South-Eastern Norway, Kongsberg, Norway\\
$^{123}$ University of Tennessee, Knoxville, Tennessee, United States\\
$^{124}$ University of the Witwatersrand, Johannesburg, South Africa\\
$^{125}$ University of Tokyo, Tokyo, Japan\\
$^{126}$ University of Tsukuba, Tsukuba, Japan\\
$^{127}$ Universit\"{a}t M\"{u}nster, Institut f\"{u}r Kernphysik, M\"{u}nster, Germany\\
$^{128}$ Universit\'{e} Clermont Auvergne, CNRS/IN2P3, LPC, Clermont-Ferrand, France\\
$^{129}$ Universit\'{e} de Lyon, CNRS/IN2P3, Institut de Physique des 2 Infinis de Lyon, Lyon, France\\
$^{130}$ Universit\'{e} de Strasbourg, CNRS, IPHC UMR 7178, F-67000 Strasbourg, France, Strasbourg, France\\
$^{131}$ Universit\'{e} Paris-Saclay, Centre d'Etudes de Saclay (CEA), IRFU, D\'{e}partment de Physique Nucl\'{e}aire (DPhN), Saclay, France\\
$^{132}$ Universit\'{e}  Paris-Saclay, CNRS/IN2P3, IJCLab, Orsay, France\\
$^{133}$ Universit\`{a} degli Studi di Foggia, Foggia, Italy\\
$^{134}$ Universit\`{a} del Piemonte Orientale, Vercelli, Italy\\
$^{135}$ Universit\`{a} di Brescia, Brescia, Italy\\
$^{136}$ Variable Energy Cyclotron Centre, Homi Bhabha National Institute, Kolkata, India\\
$^{137}$ Warsaw University of Technology, Warsaw, Poland\\
$^{138}$ Wayne State University, Detroit, Michigan, United States\\
$^{139}$ Yale University, New Haven, Connecticut, United States\\
$^{140}$ Yonsei University, Seoul, Republic of Korea\\
$^{141}$  Zentrum  f\"{u}r Technologie und Transfer (ZTT), Worms, Germany\\
$^{142}$ Affiliated with an institute covered by a cooperation agreement with CERN\\
$^{143}$ Affiliated with an international laboratory covered by a cooperation agreement with CERN.\\

\end{flushleft} 

\end{document}